\documentclass{naturep}
\usepackage[utf8]{inputenc}
\usepackage{lineno}
\usepackage{setspace}
\usepackage{color}
\usepackage{graphicx}
\usepackage{booktabs}
\usepackage{amssymb}
\usepackage{amsmath}
\usepackage{hyperref}
\usepackage{adjustbox}
\usepackage{subcaption}
\usepackage{tablefootnote}
\usepackage[margin=0.6in]{geometry}
\usepackage{verbatim}
\usepackage{caption}

\newcommand*{\tabindent}{\hspace{5mm}}

\title{\raggedright{\textbf{Weakly Supervised AI for Efficient Analysis of 3D Pathology Samples}}}
\raggedright{\author{Andrew H. Song$^{1,2,3,4}$, Mane Williams$^{5}$, Drew F. K. Williamson$^{1,2,3,4}$, Guillaume Jaume$^{1,2,3,4}$, Andrew Zhang$^{6}$, Bowen Chen$^{1,2}$, Robert Serafin$^{7}$, Jonathan T. C. Liu$^{7}$, Alex Baras$^{8,9}$, Anil V. Parwani$^{10}$, and Faisal Mahmood$^{1,2,3,4, \ast}$}}
\date{}
    
\makeatletter
\let\saved@includegraphics\includegraphics
\AtBeginDocument{\let\includegraphics\saved@includegraphics}
\renewenvironment*{figure}{\@float{figure}}{\end@float}
\makeatother

\begin{document}
\maketitle
\begin{affiliations}
 \item Department of Pathology, Brigham and Women's Hospital, Harvard Medical School, Boston, MA, USA
 \item Department of Pathology, Massachusetts General Hospital, Harvard Medical School, Boston, MA, USA
 \item Cancer Program, Broad Institute of Harvard and MIT, Cambridge, MA, USA
 \item Data Science Program, Dana-Farber Cancer Institute, Boston, MA, USA
 \item Department of Biomedical Informatics, Harvard Medical School, Boston, MA, USA
 \item Harvard-MIT Division of Health Sciences and Technology, Massachusetts Institute of Technology, Cambridge, MA, USA
 \item Department of Mechanical Engineering, Bioengineering, and Laboratory Medicine \& Pathology, University of Washington, Seattle, WA, USA
 \item Department of Pathology, Johns Hopkins Hospital, Baltimore, MD, USA
  \item Department of Biomedical Engineering, Johns Hopkins Hospital, Baltimore, MD, USA
 \item Department of Pathology, The Ohio State University, Columbus, Ohio, USA
 \item[$^{\ast}$] \textbf{Corresponding author}: Faisal Mahmood (FaisalMahmood@bwh.harvard.edu)
 \end{affiliations}

\begin{abstract}
    Human tissue consists of complex structures that display a diversity of morphologies, forming a tissue microenvironment that is, by nature, three-dimensional (3D).
    However, the current standard-of-care involves slicing 3D tissue specimens into two-dimensional (2D) sections and selecting a few for microscopic evaluation\cite{farahani2015whole, liu2021harnessing}, with concomitant risks of sampling bias and misdiagnosis\cite{king2000prostate, mehra2011impact, olson2011frozen, kim2013pathologic}. 
    To this end, there have been intense efforts to capture 3D tissue morphology and transition to 3D pathology, with the development of multiple high-resolution 3D imaging   modalities\cite{roberts2012toward, kiemen2022coda, kiemen2023tissue, huisken2004optical, tanaka2017whole, glaser2017light, glaser2019otls, glaser2022hybrid,  ritman2004micro, katsamenis2019x, lin2022emerging, tu2016stain}. However, these tools have had little translation to clinical practice as manual evaluation of such large data by pathologists is impractical and there is a lack of computational platforms that can efficiently process the 3D images and provide patient-level clinical insights. 
    Here we present Modality-Agnostic Multiple instance learning for volumetric Block Analysis (MAMBA), a deep-learning-based platform for processing 3D tissue images from diverse imaging modalities and predicting patient outcomes. 
    Archived prostate cancer specimens were imaged with open-top light-sheet microscopy\cite{glaser2017light, glaser2019otls, glaser2022hybrid} or microcomputed tomography\cite{ritman2004micro, katsamenis2019x} and the resulting 3D datasets were used to train risk-stratification networks based on 5-year biochemical recurrence outcomes via MAMBA. With the 3D block-based approach, MAMBA achieves an area under the receiver operating characteristic curve (AUC) of 0.86 and 0.74, superior to 2D traditional single-slice-based prognostication (AUC of 0.79 and 0.57), suggesting superior prognostication with 3D morphological features. Further analyses reveal that the incorporation of greater tissue volume improves prognostic performance and mitigates risk prediction variability from sampling bias, suggesting that there is value in capturing larger extents of spatially heterogeneous 3D morphology.
    With the rapid growth and adoption of 3D spatial biology and pathology techniques by researchers and clinicians, MAMBA provides a general and efficient framework for 3D weakly supervised learning for clinical decision support and can help to reveal novel 3D morphological biomarkers for prognosis and therapeutic response.
    
\end{abstract}

\textbf{Interactive Demo:} \url{https://mamba-demo.github.io/demo/}

\newpage

\begin{spacing}{1.5}
Human tissues are collections of diverse heterogeneous structures that are intrinsically three-dimensional (3D). Despite this, the examination of thin two-dimensional (2D) tissue sections mounted on glass slides has been the diagnostic standard for over a century. 2D tissue sampling represents only a small fraction of the complex morphological information inherent in 3D\cite{farahani2015whole, liu2021harnessing}. Indeed, it has been shown that diagnoses are more accurate when multiple levels are examined from the same tissue block instead of a single 2D slice \cite{king2000prostate, mehra2011impact, olson2011frozen, kim2013pathologic}.
These elements suggest that a shift from 2D to 3D pathology may allow better characterization of the morphological diversity within an entire tissue volume\cite{tanaka2017whole} and ultimately improve patient diagnosis, prognosis, and prediction of treatment response.
To holistically capture volumetric tissue morphologies, a plethora of 3D imaging techniques have emerged over the past decade\cite{roberts2012toward, kiemen2022coda, kiemen2023tissue, huisken2004optical, tanaka2017whole, glaser2017light, glaser2019otls, glaser2022hybrid, lin2022emerging, tu2016stain,  ritman2004micro, katsamenis2019x}. 
In addition to protocols for serial sectioning of tissue followed by 3D reconstruction\cite{roberts2012toward, kiemen2022coda, kiemen2023tissue}, non-destructive imaging modalities such as high-throughput 3D light-sheet microscopy\cite{huisken2004optical, tanaka2017whole, glaser2017light, glaser2019otls, glaser2022hybrid}, microcomputed tomography (microCT)\cite{ritman2004micro, katsamenis2019x}, photoacoustic microscopy\cite{lin2022emerging}, and multiphoton microscopy\cite{tu2016stain} have shown potential for capturing high-resolution 3D volumetric tissue images. However, significant barriers to the clinical adoption of 3D imaging techniques still exist. First is the challenge of analyzing the large, feature-rich 3D datasets that these techniques routinely generate: the addition of the depth dimension can increase the size of high-resolution histology images by several orders of magnitude and renders manual examination of tissue by pathologists, a workflow that can already be tedious in 2D, even more time-consuming and error-prone without assistance in analyzing the data. 
Second, with pathologists trained only on 2D slices of histopathological entities, the interpretation of 3D morphologies will likely pose additional challenges, compounded by the different visual idiosyncracies of each imaging modality.

In addressing these challenges, a computational approach such as one based on deep learning (DL)\cite{lecun2015deep, mckinney2020international, lu2021ai, he2023blinded} presents an attractive solution, as it can provide diagnostic determinations and decision support efficiently and automatically. While DL-based computational pathology frameworks have made breathtaking advances, especially for patient-level clinical predictions with minimal clinician interventions, they are nearly exclusively based on 2D tissue images\cite{campanella2019clinical, van2021deep, lu2021data, bulten2022artificial}. In 3D computational pathology, recent computational prognostication works utilize hand-engineered features derived from glands in 3D\cite{xie2022prostate} or cell nuclei\cite{serafin2023non}. The algorithms are based on predefined morphometric descriptors that are limited in scope and require sophisticated networks trained on extensive annotations to delineate these entities. 
Even in the broader 3D biomedical imaging community, research efforts have largely relied on the identification and segmentation of morphologies, requiring annotations at pixel-level\cite{taleb20203d, stringer2021cellpose, tang2022self} or slice-level annotations\cite{ouyang2020video, huang2020penet, lotter2021robust}.

To address these limitations, we introduce \textbf{M}odality-\textbf{A}gnostic \textbf{M}ultiple instance learning for volumetric \textbf{B}lock \textbf{A}nalysis (MAMBA). MAMBA is a DL-based computational pipeline for volumetric image analysis that can perform weakly-supervised patient prognostication based on 3D morphological features without the need for manual annotations by pathologists (\textbf{Figure \ref{fig:mamba}}). The defining hallmark of MAMBA is its utility as a general-purpose computational tool for tissue volume analysis. MAMBA is agnostic towards imaging modality, especially important in the current landscape of diverse 3D tissue imaging modalities\cite{roberts2012toward, kiemen2022coda, kiemen2023tissue, huisken2004optical, tanaka2017whole, glaser2017light, glaser2019otls, glaser2022hybrid, lin2022emerging, tu2016stain,  ritman2004micro, katsamenis2019x}. Furthermore, as a 3D volume can be treated as a stack of 2D planes, MAMBA enables either 2D or 3D processing of volumetric data and allows direct comparison between these approaches. 

\section*{Computational platform for weakly supervised analysis of 3D pathology samples}
\begin{figure}[!ht]
    \centering
    \includegraphics[width=0.95\textwidth]{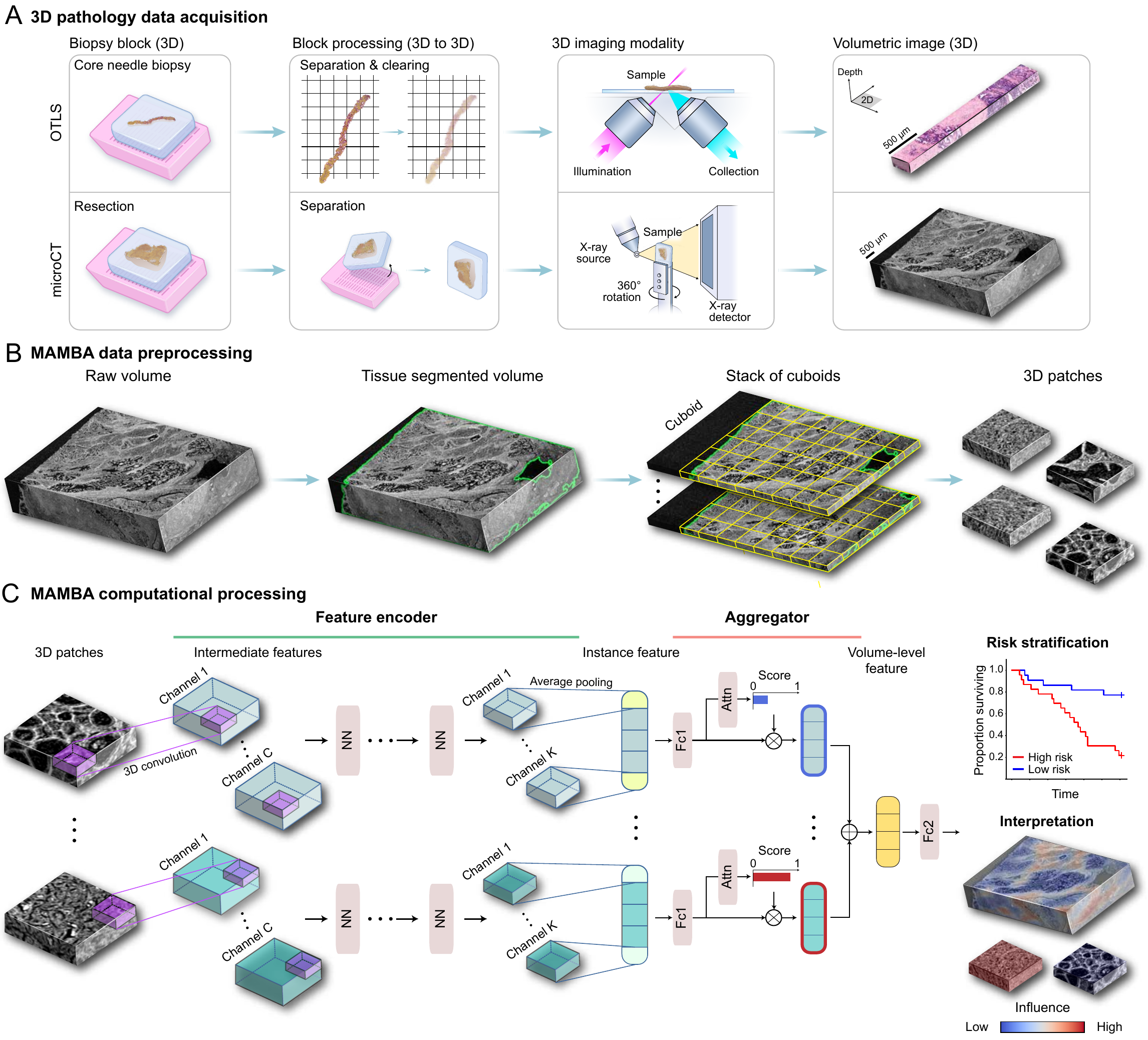}
    \caption{\textbf{MAMBA computational workflow} (A) With 3D imaging modalities such as open-top light-sheet microscopy (OTLS) and microcomputed tomography (microCT), high-resolution volumetric images of tissue specimens are captured. (B) MAMBA accepts raw volumetric tissue images from diverse imaging modalities as inputs. MAMBA first segments the volumetric image to separate tissue from the background. In a common version of the workflow, the segmented volume is then treated as a stack of cuboids (3D planes) and further tessellated into smaller 3D patches. (C) The patches (\textit{i.e.,} instances) are then processed with a pretrained feature encoder network of choice, leveraging transfer learning to produce a set of compact and representative features. The encoded features are further compressed with a domain-adapted shallow, fully-connected network. Next, an aggregator module aggregates the set of instance features, automatically weighing them according to the importance towards rendering the prediction to form a volume-level feature. MAMBA also provides saliency heatmaps for clinical interpretation and validation. The computational workflow of MAMBA with 2D processing is identical. Further details of the model architecture are described in the Methods. NN, generic neural network layers dependent on the choice of feature encoder; Channel C, K, intermediate channels in feature encoder; Attn, attention module; Fc1, Fc2, fully-connected layers.}
    \label{fig:mamba}
\end{figure}

MAMBA is comprised of three steps that distill a very large ($>10^9$ voxels) input volume into a single low-dimensional feature vector (length of $\sim10^3$) representing the input. This single vector generates a patient-level prediction for a clinical endpoint, a typical paradigm for multiple instance learning (or weakly-supervised) problems\cite{ilse2018attention}. First, the preprocessing component segments the volume into a stack of planes (2D) or cuboids (3D) that contain tissue and further tessellates them into smaller 2D or 3D patches (\textit{i.e.}, instances), which allows direct computational processing (\textbf{Figure \ref{fig:mamba}B}). Each instance is further compressed with a sequence of a pretrained 2D or 3D DL-based feature encoder network and a task-adaptable shallow network. Finally, the set of instance features from the volume is weighed and aggregated to form a volume-level feature for patient-level risk prediction. MAMBA uses an attention-based aggregation module\cite{ilse2018attention, lu2021data} to automatically identify important instances and regions contributing to prognostic decisions without additional pathologist manual annotations. As a post hoc interpretation method, saliency heatmaps for the network prediction can be used to help understand morphological correlates of the outcome labels (\textbf{Figure \ref{fig:mamba}C}).  

Volume-based 3D analysis opens new avenues for pathologists and researchers not possible with current 2D frameworks. From a clinical perspective, MAMBA can reliably include prognostically important regions not present in the traditional WSIs, which have limited coverage of morphologically heterogeneous tissue. From a technical perspective, in addition to 2D-based architectures pretrained on 2D natural images, MAMBA adopts 3D convolutional neural networks (CNN) or 3D vision transformers (VIT) pretrained on image sequences to encode 3D-morphology-aware low-dimensional features from 3D patches, which are better suited than 2D alternatives\cite{kamnitsas2017efficient, taleb20203d, esteva2021deep}. The automatic encoding of representation with a DL-based feature encoder\cite{lecun2015deep} obviates the need for hand-engineered features that are limited by human cognition along with sophisticated segmentation networks to delineate these entities\cite{stringer2021cellpose, xie2022prostate, serafin2023non}, both of which are especially challenging in 3D due to lack of annotations and differences in visual textures. Besides differences in performance, 3D processing is computationally more efficient once the patch features are extracted - The number of 3D patches and, thus, the instance features is significantly smaller than the 2D counterparts for the same volume, leading to a reduction of memory and time required for network training and testing. 

We first evaluate MAMBA on a classification task with simulated 3D phantom datasets followed by prognostication tasks for two different prostate cancer cohorts\cite{litwin2017diagnosis} imaged with different modalities. We extensively compare several analytical treatments of the volumetric samples, from using 2D patches from a single plane within each volume (emulating a traditional 2D pathology workflow) to using 3D patches from the whole volume.  

The prostate cancer datasets are curated from two different centers, Brigham and Women's Hospital (BWH) and University of Washington (UW) (BWH cohort: $n=45$ prostatectomy specimens, UW cohort: $n=50$ simulated core needle biopsies extracted from prostatectomy specimens) (Extended Data Table \ref{table:data_summary}). Gigavoxel volumetric images were used from two 3D pathology imaging modalities: open-top light-sheet microscopy (OTLS)\cite{glaser2017light, glaser2019otls, glaser2022hybrid} for the UW cohort and microcomputed tomography (microCT)\cite{ritman2004micro, katsamenis2019x} for the BWH cohort. OTLS is a fluorescence microscopy modality that allows rapid and high-resolution imaging of deparaffinized tissue volumes stained with a fast small-molecule fluorescent analog of H\&E staining and then optically cleared with a reversible dehydration-based protocol\cite{xie2022prostate}. MicroCT reconstructs a 3D volume from X-ray projections penetrating the sample embedded in paraffin from a multitude of angles. For our study, we use dual-channel (nuclear and eosin channel) OTLS at $1\mu m$/voxel resolution and single-channel (grayscale) microCT at $4\mu m$/voxel resolution. We choose prostate cancer to validate MAMBA since the important glandular and architectural features for prostate cancer prognosis\cite{bulten2020automated} can reliably be captured at varying spatial resolutions. For the patient-level clinical endpoint, we used the elapsed time from prostatectomy to prostate cancer recurrence as determined by the rise of prostate-specific antigen (PSA) above a certain threshold, termed biochemical recurrence (BCR). Further details on the patient cohort can be found in Extended Data Table \ref{table:data_summary} and the Methods.

\section*{Validation with simulated 3D data}
Following the standard practice of utilizing simulated phantom datasets to evaluate the processing pipeline and test specific data or network-related hypotheses\cite{shepp1974fourier, marabini19983d, Svoboda2009generation}, MAMBA provides functionality for users to create 3D phantom datasets. In this study, we create phantom datasets populated with distinctive 3D cells for classification and survival prediction tasks, with 3D cells of different eccentricity distributions characterizing different subtypes or risk groups (\textbf{Extended Data Figure \ref{fig:simulation}A}). We test whether performance is affected by different treatments of volumetric data as 2D or 3D entities or by sampling a portion of the volumes.

Our results suggest that utilization of the entire volume is better than only using a portion as demonstrated in the area under the receiver operating curve (AUC) metric comparison (Whole volume cuboid AUC: $0.974$, whole volume plane AUC $0.803$ vs. single plane AUC $0.677$) (\textbf{Extended Data Figure \ref{fig:simulation}B}, \textbf{Extended Data Figure \ref{fig:clf_ablation}A}). For the binary classification task, the targeted plane approach (where a targeted plane is chosen for each volume, ensuring morphologies of both classes are captured) outperforms the random plane approach (AUC $0.677$ vs. $0.501$) and demonstrates how randomly sampling a slice can miss relevant phenotypes and affect the performance\cite{liu2021harnessing}. In addition, 3D patching outperforms its 2D counterpart (AUC$=0.974$ vs. AUC$=0.803$), suggesting that 3D-morphology-aware feature encoding can significantly improve performance. The principal component feature plot for the samples with the whole volume cuboids approach also shows a good separation between the two classes as expected (\textbf{Extended Data Figure \ref{fig:simulation}C}). We also create a 3D phantom dataset for survival prediction ($n=150$) with two morphological classes assigned to two different risk groups\cite{goldstein2020x}. We observe similar trends as in the classification task, where better patient stratification is achieved with the whole volume cuboid approach (\textbf{Extended Data Figure \ref{fig:simulation}D}). Further details of the simulation datasets are explained in the Methods.

\begin{figure}[!ht]
    \centering
    \includegraphics[width=\textwidth]{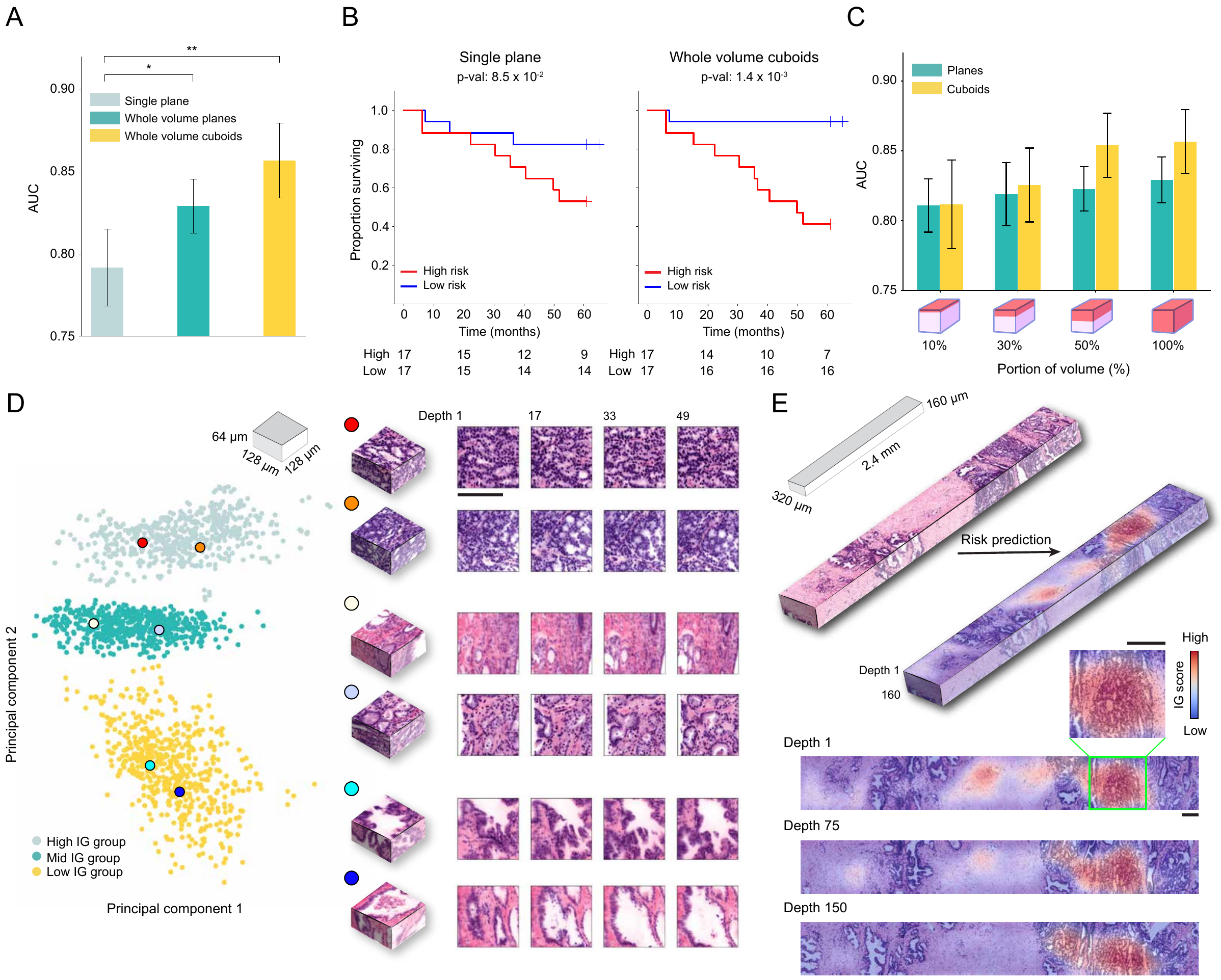}
    \caption{\textbf{MAMBA analysis of open-top light-sheet microscopy (OTLS) prostate cancer cohort.} The OTLS cohort contains volumetric tissue images ($1 \mu m$/voxel resolution) of simulated core needle biopsies extracted from prostatectomy specimens. (A) Cohort-level AUC for MAMBA trained and tested on the top plane from each volume (single plane), all planes and cuboids within the whole volume (whole volume planes and whole volume cuboids, respectively), repeated over five different experiments. Results for balanced accuracy and F1-score metrics can be found in \textbf{Extended Data Figure \ref{fig:clf_ablation}B}. Statistical significance was assessed with an unpaired t-test. $^{*}P\leq 0.05$ and $^{**}P\leq 0.01$. (B) Kaplan-Meier survival analysis for patients with BCR timestamps available, stratified at 50 percentile based on MAMBA-predicted risk, for single plane and whole volume cuboids approaches. The log-rank test was used. (C) Ablation analysis with training and testing on increasing portions from the top of each volume. (D) Principal component feature space plot for 3D patches with high (unfavorable outcome), middle (no influence), and low (favorable outcome) 10\% integrated gradient (IG) scores aggregated across the entire cohort. Representative 3D patches and 2D slices within the patch are displayed for each cluster. (E) 3D IG heatmap with representative 2D planes displaying unfavorable (red) and favorable (blue) prognostic regions. Additional examples of IG heatmaps can be found in \textbf{Extended Data Figure~\ref{fig:otls_heatmaps}}. All scale bars are $100 \mu m$.}
    \label{fig:otls}
\end{figure}

\section*{Evaluation on the OTLS cohort}
Following favorable results for the whole volume cuboid approach on the phantom datasets, we test MAMBA on the OTLS cohort for a risk stratification task. This task is cast as a binary classification task between low and high-risk groups\cite{xie2022prostate, foersch2023multistain}, with high-risk group defined as patients experiencing BCR within five years of prostatectomy. Due to the limited sample size, we perform 5-fold cross-validation with 80\% and 20\% splits between train and test data and we aggregate the predicted probability values over all folds to compute cohort-level AUCs, repeating the analysis over five different data splits. 
Similar to the simulation analysis, we compare the performance of the three prognostic models trained on 2D patches extracted from the top plane of each volume, all planes of each volume, and finally, 3D patches extracted from each volume (\textbf{Figure \ref{fig:otls}A}, \textbf{Extended Data Figure \ref{fig:clf_ablation}B}). We observe that there is a statistically significant difference between using the top plane of the volume (AUC $0.792$) and the whole volume (planes AUC $0.829$ $P<0.05$, cuboids AUC $0.857$ $P<0.01$), also supported by better cohort stratification into high and low-risk groups (\textbf{Figure \ref{fig:otls}B}). We further conduct an ablation analysis to determine whether the percentage of each volume used for training and testing affects the performance.  This is done by gradually increasing the volume seen by the model extending downward from the top of each volume. We identify an upward AUC trend as larger portions of the volumes are incorporated, regardless of whether they are processed as a stack of planes or cuboids (\textbf{Figure \ref{fig:otls}C}).  

To investigate phenotypes driving the network's risk prediction, we employ integrated gradient (IG) interpretability analysis\cite{sundararajan2017axiomatic, lee2022derivation}, where an IG score is assigned for each 2D or 3D patch upon prediction, with positive (high) scores for regions associated with increasing the predicted risk (unfavorable prognosis). The negative (low) scores are associated with regions that decrease the predicted risk (favorable prognosis).
To first understand broad morphological descriptors of favorable and unfavorable prognosis, we assemble IG values for all 3D patches within the cohort and examine patches within high (top 10 percentile), medium (10 percentile centered at 0 - agnostic towards prognostication), and low (bottom 10 percentile) IG groups (\textbf{Figure \ref{fig:otls}D}, \textbf{Extended Data Figure \ref{fig:otls_ig_ablation}A}). 
Patches from the high IG cluster exhibit infiltrative carcinoma that resembles predominantly poorly-differentiated glands (Gleason pattern 4), exhibiting cribriform architecture. Patches from the middle IG cluster exhibit infiltrative carcinoma that resembles mixtures of Gleason patterns 3 and 4. Patches from the low IG cluster predominantly exhibit large, benign glands, with occasional corpora amylacaea.
These observations concur with well-known prognostic biomarkers for prostate cancer\cite{litwin2017diagnosis}.
The IG groups form distinctive clusters in the principal component feature space, further supporting the morphological observations (\textbf{Figure \ref{fig:otls}D}) and thus indicating MAMBA.
The IG values can be overlaid on the raw volume input to generate IG interpretability heatmaps and to further locate regions of different prognostic information within the tissue volume (\textbf{Figure \ref{fig:otls}E}, \textbf{Extended Data Figure \ref{fig:otls_heatmaps}}, \textbf{Interactive Demo}).


We next check the relationship between the mean IG score of each patient to the risk predicted by the network, where the risk is defined by the predicted probability for the high-risk group (\textbf{Extended Data Figure \ref{fig:otls_ig_ablation}B}). The strong correlation (Pearson r: $0.94$, $P<0.0001$) confirms that MAMBA learns to assign prognostic attribution across the cohort. Further analysis of the IG group constituents for each sample reveals that the higher the predicted risk profile, the larger the proportion of high IG group patches and the smaller the proportion of low IG group patches (High IG Pearson r: $0.77$, $P<0.0001$, Low IG Pearson r: $-0.78$, $P<0.0001$) (\textbf{Extended Data Figure \ref{fig:otls_ig_ablation}C}). We also observe that the ratio of the number of patches in high IG and low IG groups stratifies the cohort reasonably well (\textbf{Extended Data Figure \ref{fig:otls_ig_ablation}D}), all of which collectively suggests that risk stratification is driven by the extent to which prognostic morphologies are manifested within each sample.

\section*{Evaluation on the microCT cohort}
\begin{figure}[!ht]
    \centering\includegraphics[width=0.92\textwidth]{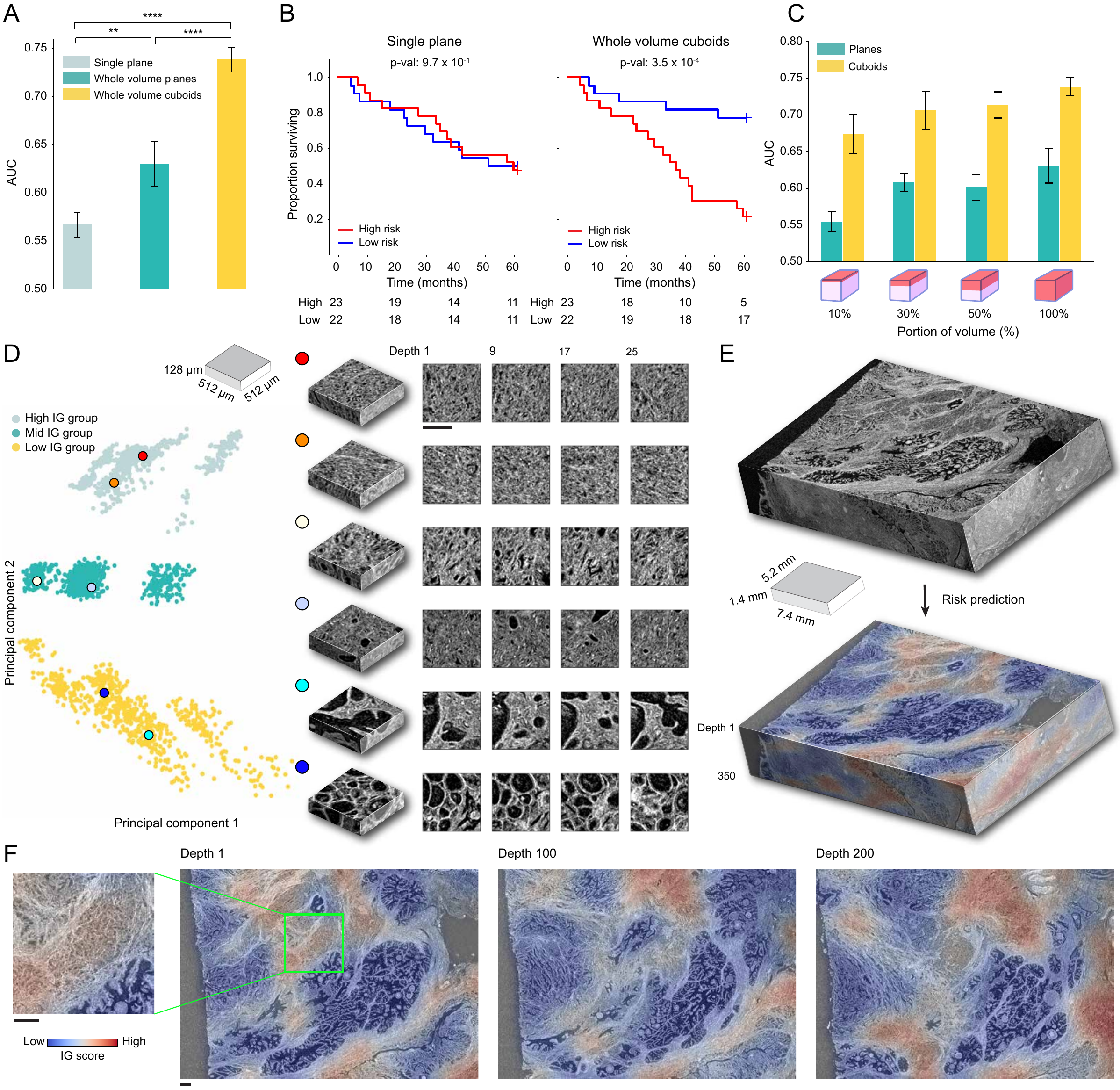}
    \caption{\textbf{MAMBA analysis of microcomputed tomography (microCT) prostate cancer cohort.} The microCT cohort contains volumetric tissue images of prostatectomy tissue from prostate cancer patients with $4 \mu m$/voxel resolution. (A) Cohort-level AUC for MAMBA trained and tested on the top plane from each volume (single plane), all planes and cuboids within the whole volume (whole volume planes and whole volume cuboids, respectively), repeated over five different experiments. Results for balanced accuracy and F1-score metrics can be found in \textbf{Extended Data Figure \ref{fig:clf_ablation}C}. Statistical significance was assessed with unpaired t-test. $^{**}P\leq 0.01$ and $^{****}P\leq 0.0001$. (B) Kaplan-Meier survival analysis, stratified at 50 percentile based on MAMBA-predicted risk, for single plane and whole volume cuboids approaches. The log-rank test was used. (C) Ablation analysis with training and testing on increasing portions from the top of each volume. (D) Principal component feature space plot for 3D patches with high (unfavorable outcome), middle (no influence), and low (favorable outcome) 10\% integrated gradient (IG) scores aggregated across the entire cohort. Representative 3D patches and 2D slices within the cuboid are displayed for each cluster. (E, F) 3D IG heatmap with representative 2D planes displaying unfavorable (red) and favorable (blue) prognostic regions. Additional heatmap examples can be found in \textbf{Extended Data Figure~\ref{fig:ct_heatmaps}}. All scale bars are $250 \mu m$.}
    \label{fig:ct}
\end{figure}

We also evaluate MAMBA on the microCT cohort for the high and low-risk stratification task, using 5-fold cross-validation with 80\% and 20\%  train and test data split, stratified by risk status repeated over five data splits. As previously observed in two datasets, the 3D feature encoding from the 3D patches yields the best performance when compared to the 2D alternatives (whole block cuboids AUC 0.739 vs. whole block planes AUC 0.631 \& single plane AUC 0.567) (\textbf{Figure \ref{fig:ct}A}, \textbf{Extended Data Figure \ref{fig:clf_ablation}C}). The superiority of the 3D approach is also reflected in better risk stratification performance (\textbf{Figure \ref{fig:ct}B}). Overall, the AUC is lower than that of OTLS, as expected, due to a lower resolution ($1 \mu m$/voxel for OTLS vs. $4 \mu m$/voxel for microCT) and single-channel information (vs. dual-channel for OTLS). Interestingly, the AUC gap between the cuboids and planes for the whole volume is statistically significant and bigger than the gap in OTLS (microCT $\Delta \text{AUC}=+0.108$, $P < 0.0001$ vs. OTLS $\Delta\text{AUC}=+0.0276$). We also observe an increasing AUC trend with the increasing percentage of the volume analyzed, consistent with OTLS (\textbf{Figure \ref{fig:ct}C}). 

Qualitative analysis of IG heatmap and representative 3D patches from each group (\textbf{Figure \ref{fig:ct}D-F}, \textbf{Extended Data Figure \ref{fig:ct_ig_ablation}A}, \textbf{Extended Data Figure \ref{fig:ct_heatmaps}}, \textbf{Interactive Demo}) display diverse morphological characteristics. The high IG cluster consists of patches with infiltrative carcinoma that most closely resembles Gleason pattern 4; however, the lower resolution and lack of H\&E staining make definitive grading effectively impossible by visual inspection of the microCT images alone. In the middle IG cluster, most patches contain infiltrating carcinoma that resembles Gleason patterns 3 and 4. The low IG cluster consists almost mostly of patches containing benign prostatic tissue with occasional foci of infiltrative carcinoma that resembles Gleason pattern 3.

Further IG score analyses for the microCT cohort agree with findings in OTLS. We observe a statistically significant correlation between the mean IG score of each patient and the predicted risk (Pearson r: $0.95$, $P<0.0001$) (\textbf{Extended Data Figure \ref{fig:ct_ig_ablation}B}). The increasing proportion of high IG patches and decreasing proportion of low IG patches within each sample with increasing predicted risk is also observed (High IG group - Pearson r: $0.79$, $P<0.0001$, Low IG group - Pearson r: $-0.61$, $P<0.0001$) (\textbf{Extended Data Figure \ref{fig:ct_ig_ablation}C}), with good cohort stratification based on the ratio of the number of patches in high IG and low IG groups (\textbf{Extended Data Figure \ref{fig:ct_ig_ablation}D}). These analyses, by and large, corroborate the OTLS dataset observation that the degree of unfavorable prognostic morphology proliferation plays an important role in risk stratification.

\section*{Cross-modal evaluation}
In the absence of independent data cohorts for validation, we instead perform cross-modal generalization experiments where a network is trained on one imaging modality with whole volume cuboids and tested on the other. This leverages that both cohorts are prostate cancer and allows for assessing whether the trained network captures generalizable morphological signatures across modalities. MAMBA trained on microCT (OTLS) data achieved an average test AUC on OTLS (microCT) cohort of 0.676 (0.725), with good separation between the high and low-risk cohorts
(\textbf{Extended Data Figure \ref{fig:cross_modality}A, B}). The drop in performance compared to training and testing on the same modality can be attributed to significantly different imaging protocols and image textures of the OTLS and microCT modalities. Despite the challenging nature of cross-modal adaptation, these results suggest that MAMBA learns to recognize important prognostic morphologies correctly and can generalize to diverse imaging modalities. Further morphological examination supports this observation, with poorly-differentiated glands or infiltrating carcinoma highlighted as contributing to unfavorable prognosis (\textbf{Extended Data Figure \ref{fig:cross_modality}C, D}).

\section*{Benefits for 3D volume analysis}
Encouraged by the consistently superior performance of volume-based prognostication compared to plane-based alternatives, we study additional benefits of the whole volume cuboid approach besides performance. To this end, we first investigate how morphological heterogeneity within a tissue volume affects risk prediction. We measure the predicted risk as a function of depth within each OTLS sample with the network trained on all planes of the tissue volume to achieve plane-level granularity. We then quantify risk fluctuation by computing the difference between the lower and upper 5\% plane-level predicted risks, effectively constructing a 90\% confidence interval. We observe that the difference gap for the cohort is wide, suggesting considerable risk variability (\textbf{Extended Data Figure \ref{fig:otls_variability_2D}A}).
Such variability can pose a challenge for a threshold-based determination of a patient risk group, with different portions of the tissue volume resulting in risk prediction on either side of the threshold (\textbf{Extended Data Figure \ref{fig:otls_variability_2D}B}) and consequently different risk groups. Further morphological analysis indeed shows that the tissue volume heterogeneity could lead to fluctuating risk, with the higher-risk plane containing significantly more Gleason grade 4 morphologies than the lower-risk slice, which is dominated by Gleason grade 3 morphologies (\textbf{Extended Data Figure \ref{fig:otls_variability_2D}C-E}).

\begin{figure}[!ht]
    \centering
    \includegraphics[width=\textwidth]{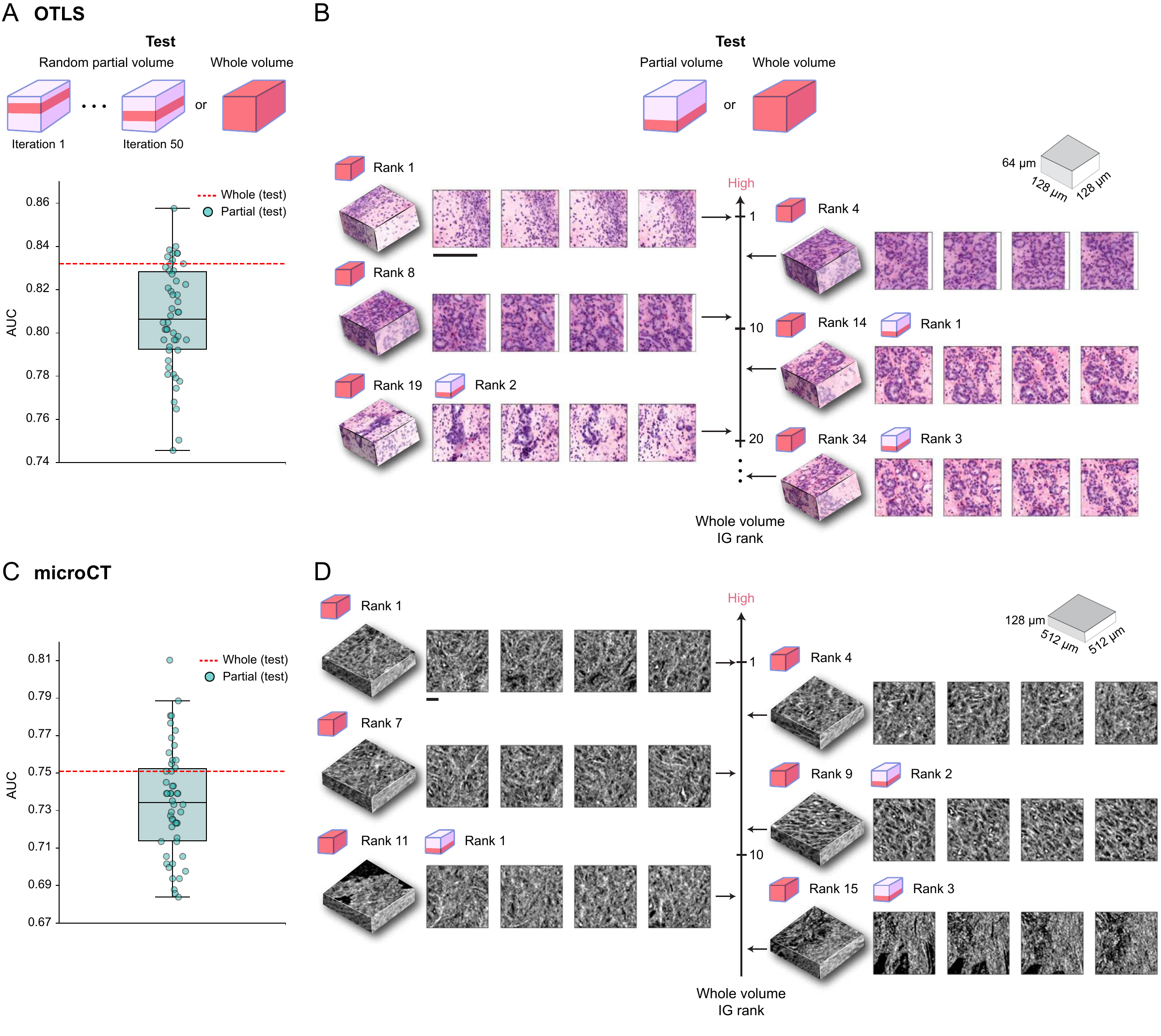}
    \caption{\textbf{Comparison between whole-volume and partial-volume analysis} Given the predictive model trained on whole volume cuboids, the cohort-level AUC is computed for the whole volume (whole volume) or over 50 iterations with 15\% of the tissue volume randomly sampled each time (partial volume). (A) OTLS cohort AUC spread for the partial volume analysis (teal) and AUC for the whole volume analysis (red). The AUC spread is considerable and indicative of significant performance variability induced by heterogeneity within the tissue volume. (B) IG score ranking for 3D patches when tested on the whole volume and partial volume of a given OTLS sample, where a higher ranking corresponds to a larger integrated gradient (IG) score. 
    The top IG score patches from the partial volume analysis are not the top contributors for increasing the risk when other patches from the whole tissue volume are accounted for. This suggests that partial volumes can miss prognostic regions. 
    (C-D) The same analyses for the microCT cohort with similar findings to the OTLS analyses. All scale bars are $100 \mu m$.}
    \label{fig:otls_variability_3D}
\end{figure}

To assess how this variability affects the cohort-level performance, we use the network trained on all cuboids from each tissue volume and test it on a portion (15\%) of each patient block, randomly sampling the portion 50 times. For the OTLS cohort, we observe that the AUC spread is considerable (median AUC: $0.806$, min-max AUC difference: $0.112$) and thus indicative of significant performance variability induced by heterogeneity within the tissue volume, with the majority (80\%) of samples falling below the AUC tested on the whole volume (\textbf{Figure \ref{fig:otls_variability_3D}A}). We observe similar results for the microCT cohort, with a comparable AUC spread (median AUC: $0.734$, min-max AUC difference: $0.126$) and the majority (74\%) of the values below AUC tested on the whole volume (\textbf{Figure \ref{fig:otls_variability_3D}C}).

\newpage

Upon careful inspection of important patches from an exemplar, we observe that the highest IG patches identified from the partial volume analysis are no longer the highest contributors to increasing the risk, when considered together with other patches from the remaining parts of the tissue in the whole volume analysis (\textbf{Figure \ref{fig:otls_variability_3D}B, D}). This suggests that important prognostic regions can often be missed when analyzing partial volumes. These results collectively reaffirm previous concerns about the role of sampling bias in diagnostic determinations\cite{king2000prostate, mehra2011impact, liu2021harnessing, olson2011frozen, kim2013pathologic}, and support the use of larger tissue volumes for reliable prognostication.

Furthermore, the whole volume cuboid approach provides better computational time and memory efficiency than 2D alternatives. While 2D and 3D patches set for a given sample require the same amount of memory, the 3D treatment results in significantly less number of patches than 2D and thus reduces the memory requirement for encoded features (\textit{i.e.}, 1D feature vectors) by roughly 30-fold (\textbf{Extended Data Table \ref{table:data}A}). 
The significantly smaller number of 3D patches also plays a crucial role in the reduction of MAMBA processing times. The feature encoding time per sample is similar for the 2D and 3D treatments (2D: 22.9 sec/sample, 3D: 26.6 sec/sample), with fewer 3D patches counteracting a much faster encoding time for a single 2D patch (2D patch: 3.2 sec/$10^4$ patches, 3D patch: 117 sec/$10^4$ patches). Since the encoded 2D and 3D instance features are of similar dimensions ($\sim 10^3$), the computational complexities are dominated by the number of patches, and therefore the network training and prediction time for 3D treatment are reduced up to 40-fold  (3D train: 0.025 sec, test: 0.011 sec vs. 2D train: 0.98 sec, test: 0.3 sec, duration computed per epoch and sample) (\textbf{Extended Data Table \ref{table:data}B}). With the imminent development of 3D imaging modalities achieving higher resolution and larger field-of-view, which will significantly expand the data size, this computational gap between 2D and 3D approaches will likely increase. Overall, the performance and computational advantages of the whole volume cuboid approach will help to alleviate these issues and facilitate the adoption of 3D approaches.

\section*{Discussion}
We present \textbf{M}odality-\textbf{A}gnostic \textbf{M}ultiple instance learning for volumetric \textbf{B}lock \textbf{A}nalysis (MAMBA), which is, to our knowledge, the first end-to-end DL-based computational platform to facilitate research on the use of 3D pathology for clinical decision support. Given a cohort of volumetric tissue images, MAMBA can seamlessly perform volumetric segmentation and 3D patching as well as combine a feature encoder of the user's choice with an attention-based aggregation network to render patient-level predictions. MAMBA additionally provides in-depth interpretability tools to probe morphological correlates that modulate patient risk. MAMBA addresses 3D tissue-based patient prognostication, a challenging task due to many factors since only a single patient-level label is provided for thousands of 3D patches without additional manual annotations (weak supervision). In addition, high data acquisition costs limit cohort size for 3D imaging datasets and each 3D imaging modality requires specific data handling procedures that are not standardized, unlike well-established staining and scanning protocols in 2D pathology.

We demonstrate the successful deployment of MAMBA for risk stratification on a simulated phantom dataset, and two prostate cancer datasets imaged with promising 3D pathology imaging modalities: open-top light-sheet microscopy\cite{glaser2017light, glaser2019otls, glaser2022hybrid} (OTLS) and microcomputed tomography\cite{ritman2004micro, katsamenis2019x} (microCT). Across the three datasets of different resolutions, textures, and volumes, we identify several trends: First, utilization of a larger portion of the tissue volume leads to better risk prediction performance, regardless of whether the data is treated as a stack of 2D images or as a 3D volume. This agrees with previous studies that have concluded that more accurate patient diagnosis and prognosis could be achieved by integrating more sections from the tissue biopsy\cite{liu2021harnessing, king2000prostate, mehra2011impact, olson2011frozen, kim2013pathologic}. Next, by treating the tissue volume as being comprised of cuboids (3D), rather than planes (2D), MAMBA achieves better patient stratification. Together, these findings suggest that 3D-morphology-aware computational frameworks, an intuitive paradigm for treating biological entities which are themselves natively 3D, can potentially improve clinical endpoint prediction.

A limitation of this study is the number of samples, despite the cohorts in the study representing unprecedented size in 3D pathology literature, due to the currently prohibitive data acquisition cost for large-scale cohort analyses. While the morphological analyses of OTLS and microCT concur with known phenotypic biomarkers (based on 2D images), a larger-scale cohort will be required to elucidate 3D biomarkers that have thus far been opaque to analysis with only 2D images. Next, to prevent overfitting with the limited sample sizes examined in this study, we did not fine-tune any of the feature encoders used for each modality, even though fine-tuning would likely have yielded better performance than transfer learning with image \& video-pretrained encoders (\textbf{Extended Data Figure \ref{fig:extractor_ablation}}). We instead trained a lightweight shallow network on top of the pretrained feature encoders to narrow the domain gap. With more data collected, further work should optimize feature encoders to each data domain and develop different aggregator modules that leverage the spatial and depth contexts within the volume. 

With decreasing costs for 3D pathology data acquisition and new protocols ensuring high-quality 3D images with resolutions at the cellular or subcellular level, recent works are beginning to demonstrate tissue heterogeneity in 3D, across diverse disease types such as pancreatic cancer\cite{kiemen2022coda}, colorectal cancer\cite{lin2023multiplexed}, melanoma\cite{merz2021high}, and kidney cancer\cite{patel2022high}. Combined with rapid advancements for AI systems in 3D computer vision, these factors will contribute to increasing demand for computational tools that can efficiently process and extract insights from volumetric tissue specimens.
MAMBA represents an important step towards satisfying this demand due to its flexibility: it is agnostic towards input modality (e.g., light-sheet microscopy, microCT) and dimensionality (2D or 3D), and the DL components, such as feature encoders, can be swapped with other appropriate architectures. Just as the introduction of computational frameworks for 2D WSI-based pathology has ushered in a boom in computational pathology research\cite{van2021deep}, MAMBA hints at the potential to assist scientists in embarking on 3D biomarker discovery endeavors and clinicians in providing better patient diagnosis and prognosis.

\section*{Online Methods}
\subsection{Patient cohorts}
The clinical end-point utilized for assessing patients' risk levels is the duration between prostatectomy and the occurrence of biochemical recurrence (BCR), marked by an elevation in prostate-specific antigen (PSA) levels surpassing a defined threshold. The precise PSA threshold triggering intervention by the treating clinician varies among clinicians and the specific laboratory conducting the PSA test, owing to variations in reference ranges across different assays. To account for this inherent variability, we deemed a patient to have reached BCR based on the date of their most recent PSA test before any intervention by the treating clinician, such as modifications in medical treatment or the initiation of radiotherapy. For both the University of Washington (UW) and Brigham and Women's Hospital (BWH) cohorts, we identified patients who had at least five years of follow-up post-radical prostatectomy (RP) with Gleason scores of 3+3, 3+4, 4+3, and 4+4 (Gleason group $1\sim 4$). For the UW cohort, archived FFPE prostatectomy specimens were collected from $n=50$ patients involved in the Canary TMA case-cohort study\cite{hawley2013biomarkers} with $n=25$ patients who experienced BCR within five years of prostatectomy and $n=25$ patients who did not. For the BWH cohort, archived FFPE prostatectomy specimens were collected from $n=64$ patients with $n=32$ patients who experienced BCR within five years of prostatectomy and $n=32$ patients who did not. Upon the microCT volumetric image quality check, $n=19$ patients were discarded due to high noise level, resulting in the cohort size of $n=45$. The detailed dataset summary can be found in \textbf{Extended Table \ref{table:data_summary}}.

\subsection{Data acquisition}
\textbf{\\Simulation data} 
We use the simulated 3D digital phantom datasets to enable comparative analyses between different network architectures and data processing approaches (e.g., 2D vs. 3D) while also ensuring every component of the MAMBA pipeline is fully functioning\cite{shepp1974fourier, marabini19983d, Svoboda2009generation}. The phantom datasets hold useful advantages over real-world clinical data in that the data is generated according to a set of pre-specified parameters, and thus all morphological characteristics that ought to be captured by the model are known already. This allows us to efficiently evaluate how well the model can capture these characteristics in the larger data regime. The user can specify different cell types, each with their own distributions for size, color, and eccentricity, as well as multiple shapes generated for each cell (\textit{e.g.}, small nuclei residing inside larger cell membranes). This is what the simulation platform of MAMBA aims to provide, with the flexibility of specifying diverse 3D morphologies and the spatial distribution of these entities within volumetric samples.

To demonstrate its utility, we design a binary class classification dataset in our study, each class populated with different distributions of normal 3D cells (represented as spheroids) and abnormal 3D cells (represented as more eccentric spheroids). Eccentricity corresponds to a mathematical concept of how \textit{stretched-out} a spheroid is. Normal cells had eccentricity drawn from a normal distribution $\mathcal{N}(0.25, 0.05^2)$, while the eccentricity of the abnormal cells followed $\mathcal{N}(0.7, 0.05^2)$. The length along the axis of symmetry (the semi-major axis) of the normal cells was drawn from $\mathcal{N}(20, (10/3)^2)$, while the length along the axis of symmetry of the abnormal cells was drawn from $\mathcal{N}(28, (10/3)^2)$. The samples of class 1 were populated with 90\% normal cells and 10\% abnormal cells. The samples of class 2 were populated with $66\%$ normal, $34\%$ abnormal cells. Finally, the thickness of the hollow spheroids was set to 3 pixels. The dimension of each sample was $512 \times 1024 \times 1024$ voxels. For each class, 50 random images were generated, each populated with 500 cells. Examples of the simulated dataset can be found in \textbf{Extended Data Figure \ref{fig:simulation}}.

In a similar context, we create a simulated survival dataset to test the cohort stratification performance of the networks. To be consistent with the OTLS and microCT dataset task, we define two risk groups defined by distinctive morphological characteristics, each with $n=75$. We use the same data generation specification as the simulated classification dataset for simplicity. The corresponding survival timepoints were synthetically generated as follows\cite{bender2005survtimes, goldstein2020x}. First, risk scores generated from $\mathcal{N}(1.0, 0.1^2)$ (class 1) and $\mathcal{N}(2.8, 0.1^2)$ (class 2) were assigned to the samples. Based on these risk scores, the survival times were generated with a Cox-Exponential model, with lower (higher) risk scores likely generating longer (shorter) survival times.
Finally, the survival times were censored by generating a cutoff point and censoring all survival times past the cutoff, where the cutoff was chosen to have approximately 30\% of the samples censored.

\paragraph{\textbf{MicroCT}}
MicroCT scanning for a series of formalin-fixed and paraffin-embedded (FFPE) cancer tissue blocks was done by Versa 620 X-ray Microscope (Carl Zeiss, Inc., Pleasanton, California, USA). Each unstained FFPE sample is attached to a plastic cassette for patient identification, which needs to be removed to avoid the plastic material absorbing X-ray and distorting the image contrast. We remove the plastic cassette by heating the entire block to partially melt the paraffin, which allows the cassette detachment with a razor blade. The separated FFPE block is mounted vertically to a custom-designed steel sample holder and placed on the stage to minimize the thermal vibration of the sample throughout the scanning. 

For each sample, two scans were performed at different resolutions. First, a quick scan at the low resolution (22.04 $\mu m$/voxel) with large field-of-view was performed to capture the whole paraffin block, followed by zooming into tumor-specific locations within the block to capture morphological details at higher resolution (3.98 $\mu m$/voxel) (Scout and Zoom protocol). For the high-resolution scan, a microfocus X-ray source with a tube voltage of 40 $kV$ and filament current of 75 $\mu A$ (3 Watts) was used. A total of 4,501 projection images, with the sample rotated 0.08 degrees (360 degrees/4,501) per projection, were captured for the entire sample on the 16-bit 3,064 pixels by 1,928 pixels flat panel detector and yielded a stack of 1,300 2D images (the depth dimension). Each projection had 15 frames for averaging and 0.5 seconds of exposure for each frame to improve the signal-to-noise ratio (a total of 7.5 seconds for each projection), with the detector recording the raw grayscale intensities for each voxel. The total scan time amounted to 11.5 hours per sample (0.5 hours for the low-resolution scan and 11 hours for the high-resolution scan), and the field-of-view $5.2\, mm \times 12.26\, mm\times 7.71\, mm $ ($1,300 \times 3,064 \times 1,928$ voxels).

All images’ grayscale intensities were first scaled using the paraffin control block (no tumor), then reconstructed with Zeiss Reconstructor software v16 (Carl Zeiss, Inc., Pleasanton, California, USA). During the scaling process, the density of the air was treated as 10 $g/cc^3$ with an average intensity of 17,030, and the density of the paraffin material (100\% wax) was treated as 27 $g/cc^3$ with an average density of 44,250. The density values were chosen to elevate the noise to well above the minimum 0 and keep the tissue material well below the possible maximum of $65,535\,(= 2^{16} -1)$ intensity value. No additional filtering operations were performed on the data.

\paragraph{\textbf{OTLS}} For each patient, FFPE tissue blocks were identified that correspond to prostate regions targeted by urologists in standard sextant and 12-core biopsy procedures. These blocks were then deparaffinized, and a roughly 1- mm diameter simulated core-needle biopsy was cut from each, resulting in each biopsy containing a tissue volume of roughly $1 \times 1 \times 15 mm$. Upon inspection, 118 biopsies that contained tumors (1 to 4 cancer-containing biopsies per patient) were selected for imaging. These cores were stained with a T\&E staining protocol\cite{xie2022prostate} (analogous to hematoxylin and eosin stain). Specifically, these biopsies were first washed with 100\% ethanol twice for 1 hour each to remove any excess xylene and then treated in 70\% ethanol for an hour to partially rehydrate them. Each biopsy was then placed in a 0.5 mL Eppendorf tube and stained for 48 hours in 70\% ethanol at pH 4 using a 1:200 dilution of Eosin-Y and a 1:500 dilution of To-PRO-3 Iodide at room temperature with gentle agitation. These biopsies were then dehydrated twice with 100\% ethanol for 2 hours. Finally, the biopsies were optically cleared by placing them in ethyl cinnamate for 8 hours.

A custom OTLS microscope\cite{glaser2019otls} was used to image each biopsy across 2 wavelength channels (with laser wavelengths of 488nm and 638nm). Ethyl cinnamate was utilized as the immersion medium, and a multi-channel laser system was used to provide the illumination. Tissues were imaged at near-Nyquist sampling of approximately $0.44 \mu m$/voxel resolution and the volumetric imaging time was approximately 0.5 minutes/$mm^3$ of tissue for each wavelength channel. For efficient computational processing, we downsample the data by a factor of 2$\times$ to ~$0.94\,\mu m /$voxel. Each volumetric image amounted to $320 \times 520 \times 9,500$ voxels. The resulting data is saved as 16-bit unsigned integers. More details on the acquisition process can be found in \cite{xie2022prostate}. 

\subsection{Volumetric image preprocessing}
\textbf{\\Volume segmentation} We treat the volumetric image as a stack of 2D images and perform tissue segmentation serially on the stack. First, the mean voxel intensity is computed for each image to identify a subset of stacks containing air and images below a user-defined threshold are disregarded prior to segmentation. Images in the remaining stack are then converted to grayscale color space, median-blurred to suppress edge artifacts, and binarized with modality-specific thresholds. The tissue contours are identified based on the binarized images, and the stack of tissue contours serves as the contour for the volume input. Images with tissue area below a certain threshold are removed to ensure sufficient tissue exists in each image.  

\noindent\textbf{3D patching \& 2D patching}
The segmented volume is patched into a set of smaller 2D patches (from a stack of planes) or 3D patches (from a stack of cuboids) to make direct computational processing of the volume feasible. The patch size and the overlap between the patches are chosen to ensure that context is sufficiently covered within each patch and enough patches exist along each dimension. For OTLS, we use 3D patch size of $128 \times 128 \times 64$ voxels ($\simeq$ $128 \times 128 \times 64\, \mu m$). An overlap of 32 voxels along the depth dimension is used to ensure that enough patches exist along the depth dimension, as it is only comprised of 320 voxels. For microCT, we use $128 \times 128 \times 32$ voxels ($\simeq$ $512 \times 512 \times 128\, \mu m$) without any overlap as the size of tissue allows a sufficient number of patches along all dimensions. For 2D patch, we use a non-overlapping patch of $128 \times 128$ pixels ($\simeq$ $128\times 128\, \mu$m for OTLS and $512\times 512\, \mu$m for microCT) for both modalities.

For 3D patching, a reference plane is required from which the patching operation along the depth dimension is started. We use the largest plane by tissue area (identified by the tissue contour from the segmentation step) as the reference and compute the two-dimensional patch coordinates within the tissue contour. We then perform 3D patching along both directions of the depth dimension starting from the reference plane. The collection of two-dimensional coordinates computed in the reference plane is used across the entire volume. Upon completion, we remove 3D patches if more than 50\% of the volume (area) constitutes the background to ensure each patch contains sufficient tissue.

After patching, the intensity in each patch is clipped at modality-specific lower and upper thresholds and then normalized to $[0, 1]$ for the next feature encoding step. For microCT, the lower threshold is set to $25,000$ intensity value and the upper threshold to the top 1$\%$ of each volume's intensity value. For OTLS, the lower threshold is set to $100$, and the upper threshold to the top 1$\%$ of each volume's intensity value. For OTLS, we additionally invert the normalized intensity values.

\subsection{Model architecture}
\textbf{\\Feature encoder choice}
Feature encoder networks serve the purpose of extracting and encoding compressed and representative descriptor $\mathbf{h}_j\in\mathbb{R}^K,\,j=1,\cdots,J$ of the patch input $\mathbf{x}_j\in\mathbb{R}^{L \times D\times H \times W}$ (3D patch) or $\mathbf{x}_j\in\mathbb{R}^{L\times H \times W}$ (2D patch), where $K$ corresponds to the encoded feature dimension, $J$ denotes the number of patches, $L$ denotes number of input channels, and $D$, $H$, $W$ denotes the depth, height, and width dimension respectively. MAMBA provides the choice between a range of 2D and 3D pretrained feature extractors based on convolutional neural networks (CNN) or vision Transformer (ViT) for transfer learning. For 3D feature encoders, MAMBA provides 3D-inflated ResNet50\cite{carreira2017quo}, spatiotemporal CNN\cite{tran2018closer, ouyang2020video}, and video sliding-window (Swin) transformer\cite{liu2022video, selva2023video}. Due to the scarcity of patient-level labels for 3D pathology datasets and generally larger encoder network size for processing the depth dimension, fine-tuning the pretrained features encoders in the volumetric image data domain results in network overfitting and poor generalization performance. To address the domain gap, we apply a fully-connected linear layer to the feature encoder outputs $\{\mathbf{h}_j \}_{j=1}^J$, parameterized by $\mathbf{W}_{\text{enc}}\in \mathbb{R}^{256\times K}$ and $\mathbf{b}_{\text{enc}}\in\mathbb{R}^{256}$, followed by GeLU nonlinearity. This further converts patch feature $\mathbf{h}_j$ from the feature encoder to a more-compressed and domain-specific feature $\mathbf{z}_j\in\mathbb{R}^{256}$ conducive to downstream tasks with better generalization performance:

\begin{equation}
    \mathbf{z}_j=\operatorname{GeLU}(\mathbf{W}_{\text{enc}}\mathbf{h}_j + \mathbf{b}_{\text{enc}})
\end{equation}

For our study, we used truncated deep residual CNN (ResNet-50) pretrained on natural images (ImageNet) for 2D experiments  and spatiotemporal CNN with ResNet-50 backbone\cite{tran2018closer, ouyang2020video} pretrained on video clips (Kinetics-400) for 3D experiments for two reasons. First, this choice ensures a fair comparison of network architecture, as both networks are based on deep residual components. Second, a further ablation study between different feature encoders shows that the spatiotemporal CNN performs the best for both OTLS and microCT datasets (\textbf{Extended Data Figure \ref{fig:extractor_ablation}}). Nevertheless, we encourage testing out different feature encoders as other encoders could yield the best performance depending on the task. 

\noindent\textbf{Feature encoding step}
As most feature encoders take three-channel RGB inputs, we emulate the setting by replicating channel information. For the dual-channel OTLS data, we replicate the nuclear channel data across the first two channels and set the eosin channel as the third. For the single-channel microCT data, we replicate the data across all three channels. For the feature encoding step, we use a batch size of 500 for 2D patches and 100 for 3D patches. The immediate output of the feature encoder is 3-dimensional for a 2D patch $(K, \widetilde{H}, \widetilde{W})$ and 4-dimensional for a 3D patch $(K, \widetilde{D}, \widetilde{H}, \widetilde{W})$, where $\widetilde{D}, \widetilde{H}, \widetilde{W}$ correspond to the downsampled depth, height, width dimension respectively. The intermediate features gets compressed to one-dimensional feature $\mathbf{h}_j\in\mathbb{R}^K$ with adaptive average-spatial pooling operation and subsequently to $\mathbf{z}_j\in\mathbb{R}^{256}$ with the fully-connected network. 

\noindent\textbf{Aggregation module}
The patching and feature encoding operation results in a collection of 256-dimensional features (also referred to as instances) $\{\mathbf{z}_j\}_{j=1}^J$, constituting the volume with a single patient-level supervisory label, a setting which is referred to as multiple instance learning (MIL). It is also referred to as weakly-supervised learning, due to the substantial size of the input (number of patches) in comparison to the supervisory label. To this end, we use an attention-based aggregation module\cite{ilse2018attention, lu2021data}, a lightweight attention network that learns to automatically compute the importance score of each patch feature and aggregates by weighted-averaging the features to form a single volume-level feature. The attention network consists of three sets of parameters $\mathbf{V}\in \mathbb{R}^{64\times 256}$, $\mathbf{U}\in \mathbb{R}^{64\times 256}$, and $\mathbf{W}\in \mathbb{R}^{1\times 64}$. The network computes assigns an importance score $a_j\in[0,1]$ to feature $\mathbf{z}_j$:

\begin{equation}
    a_j = \dfrac{\operatorname{exp}(\mathbf{W}(\operatorname{tanh}(\mathbf{V}\mathbf{z}_j)\odot \operatorname{sigm}(\mathbf{U}\mathbf{z}_j)))}{\sum_{j'=1}^J\operatorname{exp}(\mathbf{W}(\operatorname{tanh}(\mathbf{V}\mathbf{z}_{j'})\odot \operatorname{sigm}(\mathbf{U}\mathbf{z}_{j'})))},
\end{equation}
with $\operatorname{tanh}$ and $\operatorname{sigm}$ denoting hyperbolic tangent and sigmoid function respectively, and $\odot$ denoting element-wise multiplication operation. A high score ($a_j$ close to 1) indicates that the corresponding patch is very relevant for sample-level risk prediction, with a low score ($a_j$ close to 0) indicating no prognostic value. Finally, the volume-level feature $\mathbf{z}_{\text{volume}}$ is computed as 

\begin{equation}
    \mathbf{z}_{\text{volume}}=\sum_{j=1}^J a_j\mathbf{z}_j \in \mathbb{R}^{256}.
\end{equation}

\noindent\textbf{Classification module}
The volume-level feature is fed into the final linear classification layer parameterized by $\mathbf{W}_{\operatorname{cls}}\in \mathbb{R}^{1\times 256}$ and bias $\mathbf{b}_{\operatorname{cls}}\in\mathbb{R}$ resulting in the probability for the high-risk group $p$:

\begin{equation}
    p = \operatorname{sigm}(\mathbf{W}_{\operatorname{cls}}\mathbf{z}_{\text{volume}} + \mathbf{b}_{\text{cls}}).
\end{equation}

\subsection{Training \& Evaluation}
We train all the networks for a fixed number of 50 epochs and the initial learning rate of $2\times 10^{-4}$ with the cosine decay scheduler. We use AdamW optimizer with default parameters of $\beta_1=0.9$ and $\beta_2=0.999$, with weight decay of $5\times 10^{-4}$. We use a mini-batch size of 1 patient sample, pooling together patches across samples if multiple tissue samples exist for the patient. We also employ gradient accumulation of 10 training samples for training stability.
For each volume, we randomly sample 50\% of the patches as a means of data augmentation to prevent overfitting and also inject diversity into training samples. Rather than sampling randomly from the entire volume, we sample 50\% of the patches per each plane or cuboid to ensure all depths are equally accounted for. In addition, we employ a heavy dropout of $p=0.5$ after each fully-connected layer. We also apply conventional data augmentation schemes to patches, such as rotation and intensity jittering on top of random sampling of the patches. We use the binary cross-entropy loss for the loss function.

\subsection{Integrated gradients interpretability analysis}
The integrated gradient (IG) method\cite{sundararajan2017axiomatic, lee2022derivation} assesses the relationship between an input to a network and the corresponding prediction. In our study, the input and the prediction correspond to the set of  patch feature $\{\mathbf{h}_j\}_{j=1}^J$ and the probability for high-risk group $p$, respectively. The IG method assigns an IG score for each input ($\mathbf{h}_j$ in our case), signifying the strength of each input's influence on the prediction, with the sign of the score indicating the direction of influence. In the context of prognosis, this can directly be translated as positive IG values increasing the risk (unfavorable prognosis) and negative IG values decreasing the risk (favorable prognosis). The IG values close to 0 have no prognostic influence.

Denoting $F$ as the sequence of the fully-connected layer for feature encoder, attention aggregation, and classification modules, \textit{i.e.,} $p=F(\{\mathbf{h}_j \}_{j=1}^J)$, $M$ as the total number of IG interpolation steps, $\mathbf{h}_{j,k}$ as the $k^{\text{th}}$ element of the feature $\mathbf{h}_j$, and finally assuming zero feature baseline, we can compute the IG score for $\mathbf{h}_j$ as

\begin{equation}
    \operatorname{IG}(\mathbf{h}_j)=\sum_{k=1}^K \mathbf{h}_{j,k} \times \frac{1}{M}\sum_{m=1}^M \frac{\partial F\left(\left\{\frac{m}{M} \cdot \mathbf{h}_{j}\right\}_{j=1}^J\right)}{\partial \mathbf{h}_{j,k}}.
\end{equation}
Once all the IG scores are computed for the patch features of a patient, we normalize the negative IG values to $[-1, 0]$ and the positive IG values to $(0, 1]$, to ensure the sign and the influence of a patch does not get flipped. 

\subsection{Runtime analysis} Since MAMBA's runtime is not affected by the modality of a dataset modality when computed per sample, we report the runtime for a single modality. For feature encoding runtime analysis, we use deep-residual CNN (ResNet50, 2D patches) and spatiotemporal CNN (3D patches) for the feature encoders. We use batch size of $5,000$ and $100$ for 2D and 3D patches respectively, with both reaching 90\% GPU memory utilization. For training and inference analysis, the runtimes are computed per epoch and sample.

\subsection{Cross-modal analysis} The OTLS and microCT dataset characteristics are vastly different and thus necessary adjustments are required for fair cross-modal experiments. To this end, we downsample the OTLS dataset by a factor of 4 (from $1 \mu m$/voxel to $4 \mu m$/voxel) and use only the nuclear channel to match with $4 \mu m$/voxel single-channel microCT dataset. Each of these adjustments results in information loss and contributes to the drop in test AUC for OTLS dataset (\textit{i.e.,} trained and tested on single-channel $4 \mu m$/voxel OTLS data vs. dual-channel $1 \mu m$/voxel OTLS data). No adjustments were made to the microCT dataset. Five models trained in a cross-validation setting on one cohort are applied to the other cohort to compute the test AUC. 

\subsection{Statistical analysis} 
Training and testing on both the microCT and OTLS cohort were performed with 5-fold cross-validation stratified by the BCR status. Upon completion of training for all five folds, the predicted probabilities from each fold are aggregated and cohort-level AUC is computed. We repeated the experiments with five different random splits of the training and testing data. For the comparison of high and low-risk survival curves, we used the log-rank test. The stratification of the high and low-risk groups was always performed at 50 percentile of the predicted probability for the high-risk group, unless specified otherwise. We used the two-sided unpaired t-test to assess the statistical significance of two individual groups. Spearman's correlation coefficient $r$ was used to assess the correlation between two quantities. Differences in the compared group were considered statistically significant when $P$ values were smaller than 0.05 ($P > 0.05$, not significant; *$P \leq 0.05$, **$P \leq 0.01$, ***$P \leq 0.001$, ****$P \leq 0.0001$).

\subsection{Visualization}
\textbf{\\False-coloring the raw input} 
For the OTLS dataset, we use a false-coloring module\cite{serafin2020falsecolor} which utilizes the physics model of the dual-channel information of the raw OTLS data to render hematoxylin and eosin appearance. While MAMBA provides style transfer capability with cycleGAN network\cite{zhu2017unpaired}, we found that the translation does not yield clean translation and is prone to introduce spurious information, especially for microCT data where only single-channel information is provided. 

\noindent\textbf{Integrated gradients heatmap}
To generate fine-grained 3D IG heatmaps, we use 3D cuboid patches with 75\% overlap in 2D plane direction and 50\% overlap along the depth dimension, to reduce blocky effects. To compute an IG score for a given region, the raw IG scores (prior to normalization) of all the patches covering the region are accumulated and divided by the number of overlapping patches. These IG scores are then normalized in the manner described in the previous section. A coolwarm colormap, with red and blue colors indicating positive and negative IG values respectively, is then applied to the normalized IG scores, which is then overlaid on the raw volumetric image with a transparency value of 0.4. The IG heatmaps are shown in \textbf{Figure \ref{fig:otls}E}, \textbf{Figure \ref{fig:ct}E}, \textbf{Extended Data Figure \ref{fig:otls_heatmaps}}, \textbf{Extended Data Figure \ref{fig:ct_heatmaps}}, and can also be visualized in our interactive demo website (\url{https://mamba-demo.github.io/demo/}).

\subsection{Computational Hardware and Software}
All volumetric images were processed on AMD® Ryzen multicore CPUs (central processing units) and a total of 6 NVIDIA GeForce RTX 3090 GPUs (graphics processing units) using our custom, publicly available MAMBA package processing pipeline implemented in Python (version 3.10.9). MAMBA uses pillow (version 9.2.0) and opencv-python (version 4.6.0) for image processing. The spatiotemporal CNN and Swin transformer feature encoders were adapted from pytorchvideo (version 0.1.5). 3D Visualization was accomplished via napari (version 0.4.16). Plots were generated in Python using matplotlib (version 3.5.2) and numpy (version 1.22.4) was used for vectorized numerical computation. Other Python libraries used to support data analysis include pandas (version 1.4.3), scipy (version 1.9.0), tensorboardX (version 2.5.1), torchvision (version 0.13), and timm (version 0.4.12). The scientific computing library scikit-learn (version 1.0.2) was used to compute various classification metrics and estimate the AUC ROC. The survival analysis was performed with lifelines (0.26.0). The interactive demo website was developed using THREE.js (version 0.152.2) and jQuery (version 3.6.0).

\section*{Data availability} The OTLS prostate cancer cohort dataset is available from the TCIA (https://www.cancerimagingarchive.net/). Restrictions apply to the availability of the microCT data, which were used with institutional permission through IRB approval for the current study, and are thus not publicly available. Please email all requests for academic use of raw and processed data to the corresponding author (and also include A.H.S. (asong@bwh.harvard.edu)). All requests will be evaluated based on institutional and departmental policies to determine whether the data requested is subject to intellectual property or patient privacy obligations. Data can only be shared for non-commercial academic purposes and will require a formal material transfer agreement.

\section*{Code availability} The code is available at \url{https://github.com/mahmoodlab/mamba}

\section*{Author Contributions}
A.H.S and F.M. conceived the study and designed the experiments. M.W. created the simulation dataset. A.H.S. and D.W. curated the BWH cohort. A.H.S. and B.C. imaged the microCT dataset. A.H.S. and M.W. developed data visualization tools. A.H.S., M.W., G.J. ran all the experiments. A.Z. developed the interactive demo. R.S., J.T.C.L. provided the OTLS dataset and guidance on the data. D.W., A.B., A.V.P. examined the morphology. A.H.S. and F.M. prepared the manuscript. All authors contributed to the writing. F.M. supervised the research.

\section*{Acknowledgements}
The authors would like to thank Greg Lin at Harvard Center for Nanoscale Systems for operating the microCT scanner for data collection; Orestis Katsamenis at the University of Southampton for helpful advice on microCT scanner for soft-tissue materials; Gan Gao for assistance with OTLS dataset; Katie Yost for assisting with the figures, T. A. Mages and T. V. Ramsey for logistical support. This work was funded in part by the Brigham and Women’s Hospital (BWH) President’s Fund, Mass General Hospital (MGH) Pathology and by the National Institute of Health (NIH) National Institute of General Medical Sciences (NIGMS) through R35GM138216. D.F.K was supported by  National Cancer Institute (NCI) Ruth L. Kirschstein National Service Award through T32CA251062. MicroCT imaging reported in this publication was supported by Harvard University, Center for Nanoscale Systems funded by the NIH through S10OD023519. J.T.C.L acknowledges funding support from the Department of Defense (DoD) Prostate Cancer Research Program (PCRP) through W81WH-18-10358 and W81XWH-20-1-0851; the NCI through R01CA268207; and the National Institute of Biomedical Imaging and Bioengineering (NIBIB) through R01EB031002. The content is solely the responsibility of the authors and does not reflect the official views of the NIH, NIGMS, NCI, DoD.

\section*{Competing Interests}
J.T.C.L. is a co-founder and board member of Alpenglow Biosciences, Inc., which has licensed the OTLS microscopy portfolio developed in his lab at the University of Washington. 
\end{spacing}

\newpage
\begin{nolinenumbers}
\section*{References}
\bibliographystyle{nature}
\bibliography{3Dref}
\end{nolinenumbers}

\newpage
\setcounter{figure}{0}
\renewcommand{\figurename}{Extended Data Figure}

\begin{figure}[!ht]
    \centering
    \includegraphics[width=\textwidth]{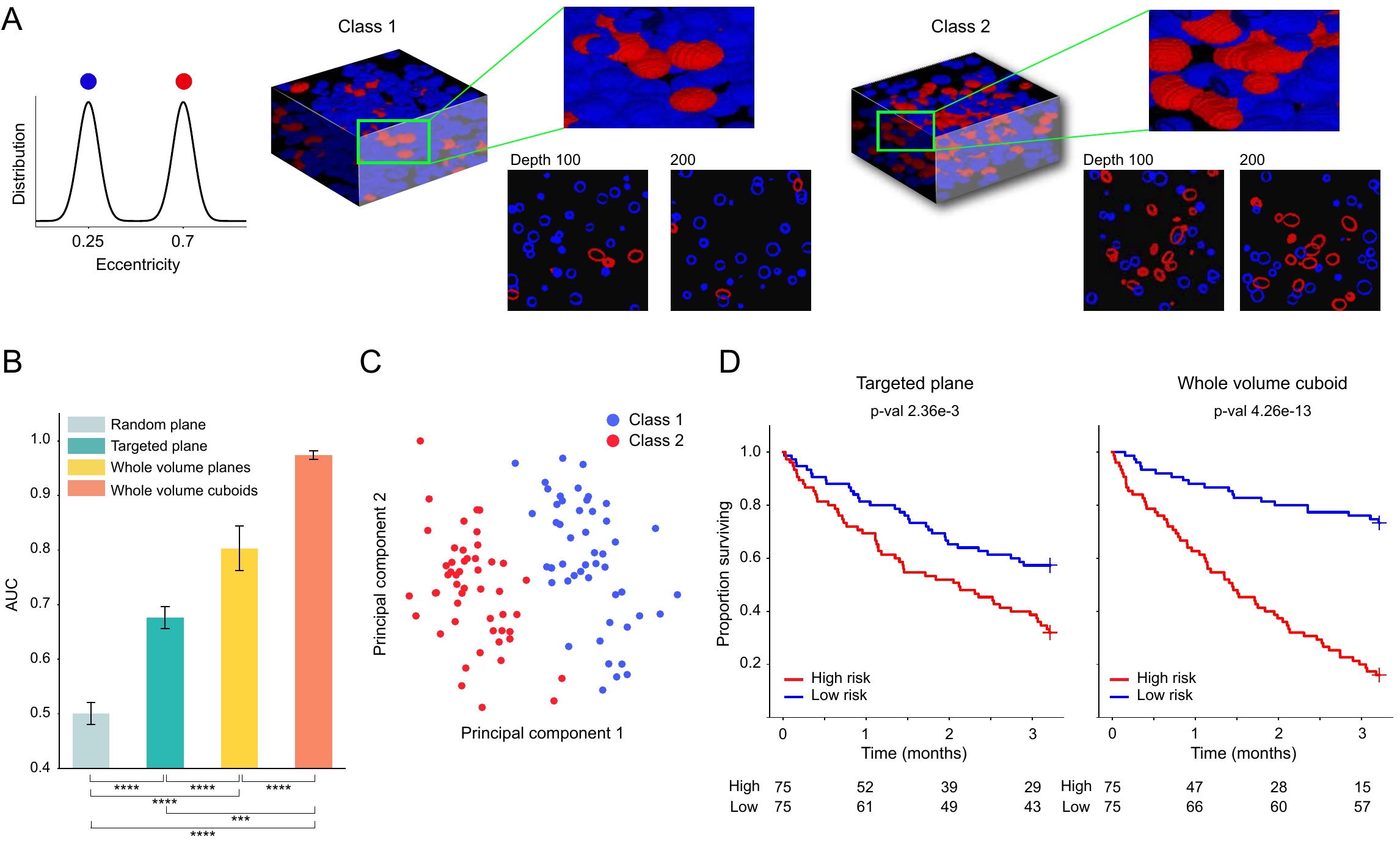}
    \caption{\textbf{3D phantom datasets and analysis with MAMBA} (A) Examples of single-channel 3D phantom data samples for the binary classification task ($n=100$), false-colored for different cell types. Samples from the first class are dominated by normal cells (blue), while samples from the second class are dominated by abnormal cells with large eccentricity (red). (B) Binary classification task AUC for MAMBA trained and tested on a random plane from each volume (random plane), the targeted plane that contains both cell types (targeted plane) from each volume, all planes, and cuboids within the whole volume (whole volume planes and cuboids). $^{***}P\leq 0.001$ and $^{****}P\leq 0.0001$. (C) Principal component feature space plot for the sample-level attention-aggregated volume features for whole volume cuboid approach, with the colors indicating ground truth labels. (D) Kaplan-Meier survival analysis stratified at 50 percentile by MAMBA-predicted risk on survival prediction phantom dataset ($n=150$) for 2D targeted single plane and whole volume cuboids approaches. Further details of the simulation dataset are described in the Methods.}
    \label{fig:simulation}
\end{figure}

\newpage
\begin{figure}[!ht]
    \centering
    \includegraphics[width=\textwidth]{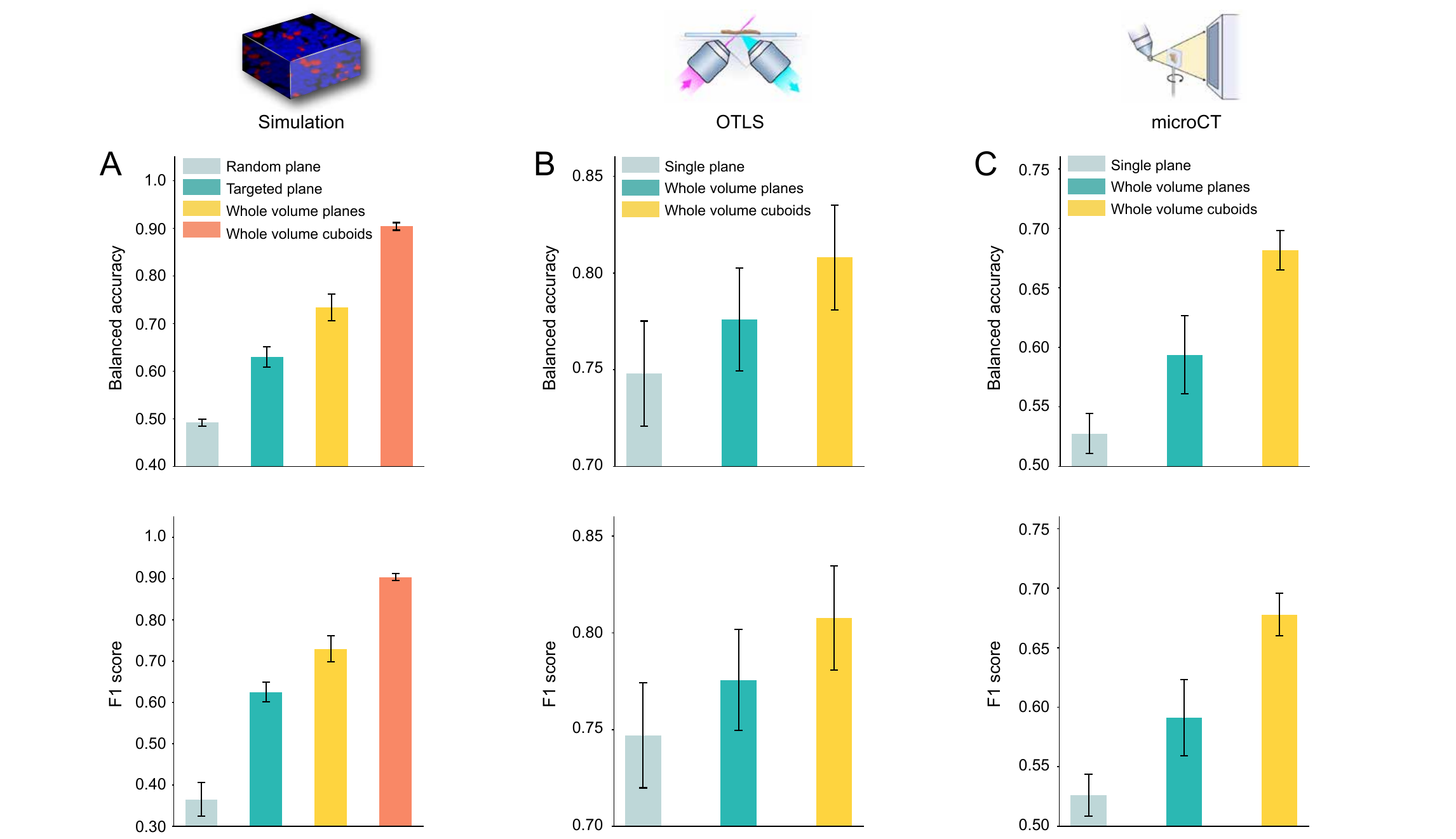}
    \caption{\textbf{Additional metrics for high \& low-risk patient classification task} Balanced accuracy and F1 score for high \& low-risk patient classification task for (A) simulation (B) OTLS (C) microCT dataset. In all three datasets and metrics, we observe that the 3D treatment of the whole volume as cuboids and 3D patching is superior to the 2D plane-based alternatives.}
    \label{fig:clf_ablation}
\end{figure}

\clearpage
\begin{figure}[!ht]
    \centering
    \includegraphics[width=0.95\textwidth]{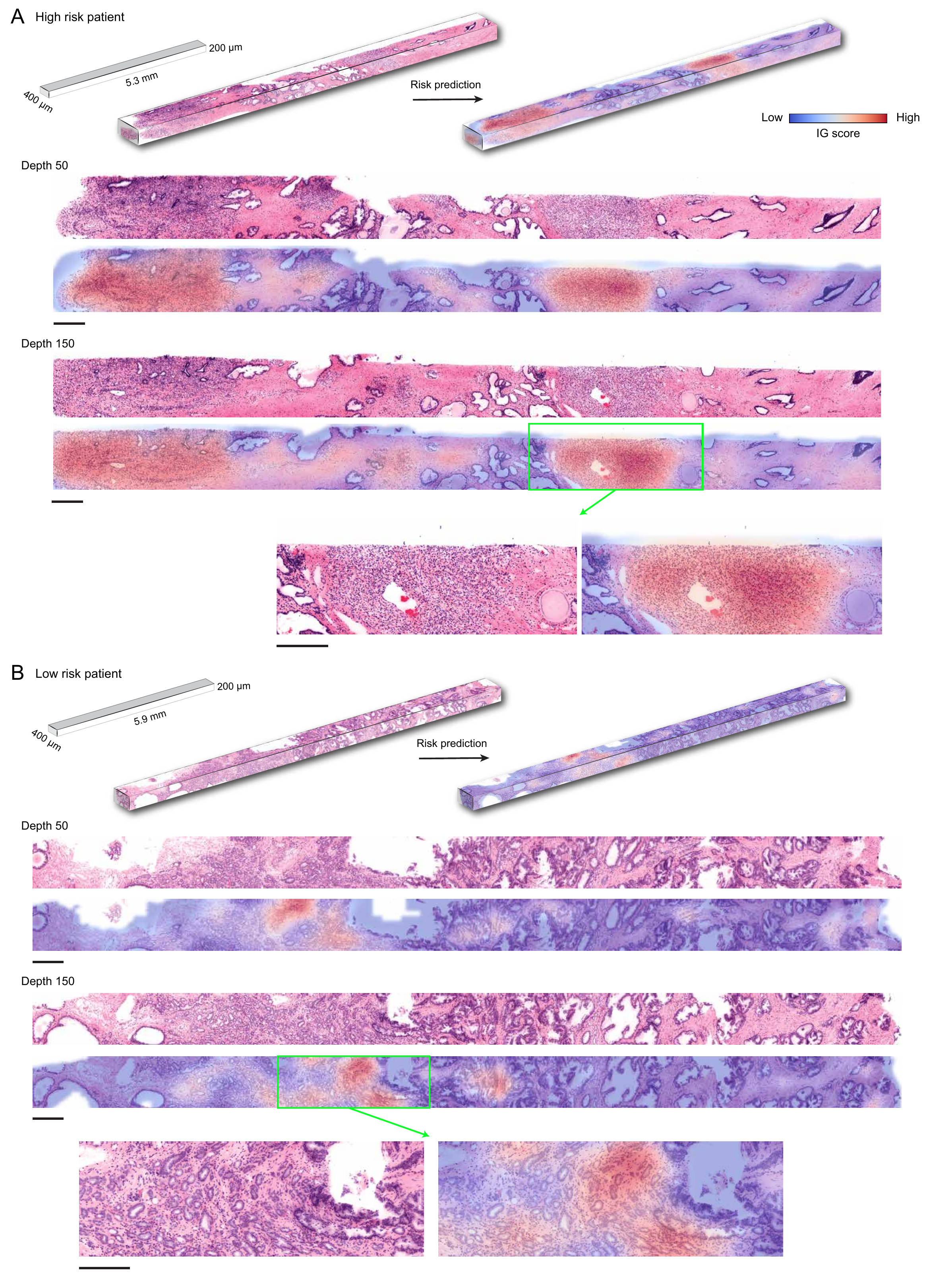}
\end{figure}

\newpage
\captionof{figure}{\textbf{Examples of integrated gradient (IG) heatmaps for open-top light-sheet microscopy (OTLS) cohort} The integrated gradient scores are assigned to each patch with high IG (low IG) patches indicating that patch contributes to an unfavorable (favorable) prognosis. The IG heatmaps are restricted to the inside of the segmented tissue contour. (A) High IG areas in the high-risk sample contain cancerous glands that resemble poorly differentiated tumor glands (Gleason patterns 4). (B) In the low-risk sample, the high IG areas are those with cancerous glands that are smaller, more tortuous, and more closely resemble Gleason pattern 4, as well as regions with a cellular stroma. All scale bars are $200 \mu m$. The heatmaps can also be visualized in our interactive demo.}
\label{fig:otls_heatmaps}

\newpage
\begin{figure}[!ht]
    \centering
    \includegraphics[width=0.93\textwidth]{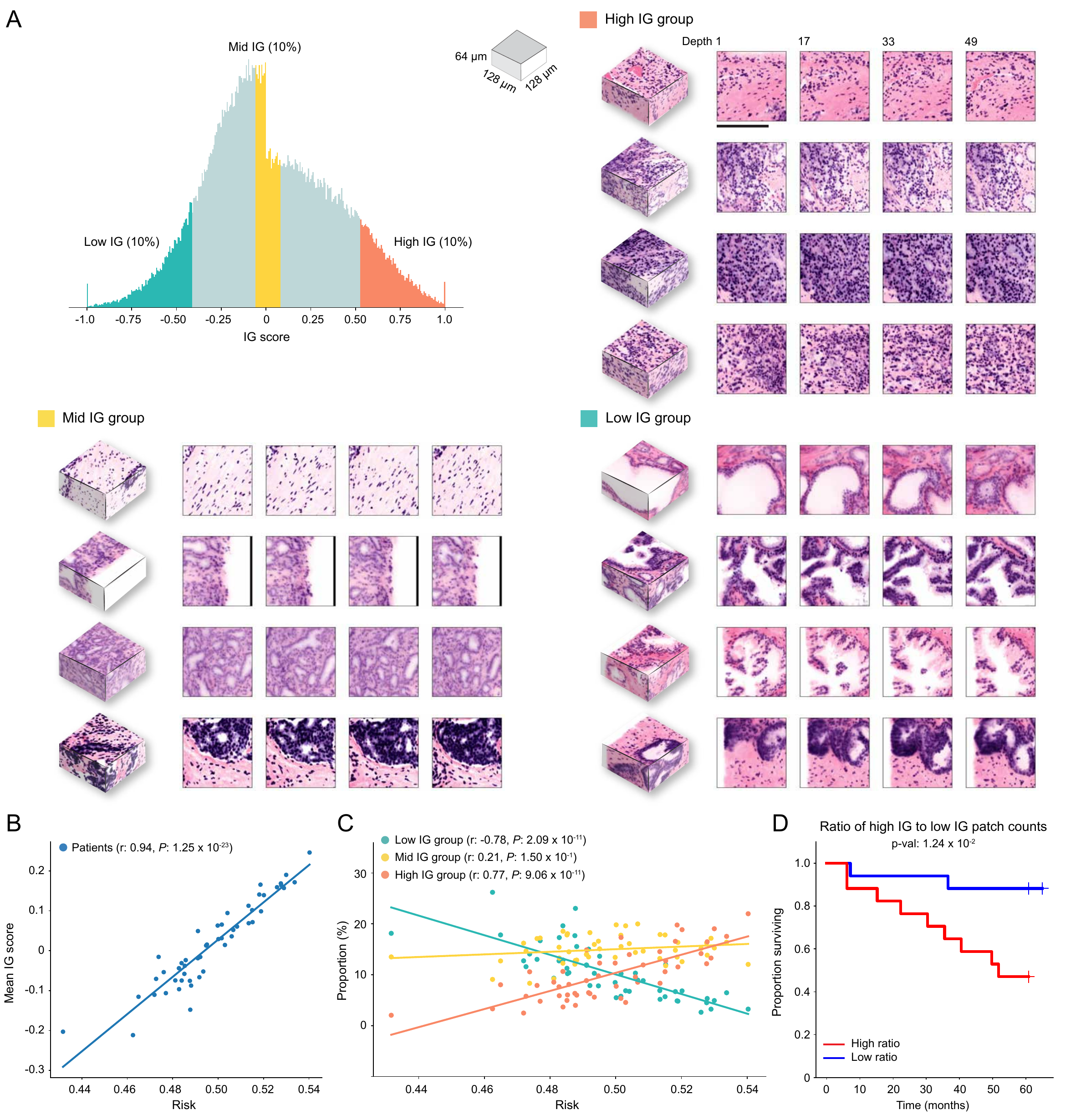}
    \caption{\textbf{Integrated gradient analysis for open-top light-sheet microscopy (OTLS) dataset} (A) Patches from the high IG cluster exhibit infiltrative carcinoma that resembles predominantly poorly-differentiated glands (Gleason pattern 4), exhibiting cribriform architecture. Patches from the middle IG cluster exhibit infiltrative carcinoma that resembles mixtures of Gleason patterns 3 and 4. 
    Patches from the low IG cluster predominantly exhibit large, benign glands, with occasional corpora amylacaea. (B) Scatter plot of the normalized IG patch scores averaged within each sample as a function of predicted risk (the predicted probability for the high-risk group). (C) The scatter plot of the proportion of the number of high, middle, and low IG group patches in each sample as a function of predicted risk, which shows that a sample with a higher predicted risk profile has a larger (smaller) fraction of high (low) IG patches. (D) Kaplan-Meier curve for the cohort stratified (50\%) by the ratio of the number of patches in the high and low IG group. The good stratification performance suggests that the extent to which prognostic morphologies manifest in each sample is also important. The scale bar is $100 \mu m$.}
    \label{fig:otls_ig_ablation}
\end{figure}

\clearpage
\begin{figure}[!ht]
    \centering
    \includegraphics[width=0.9\textwidth]{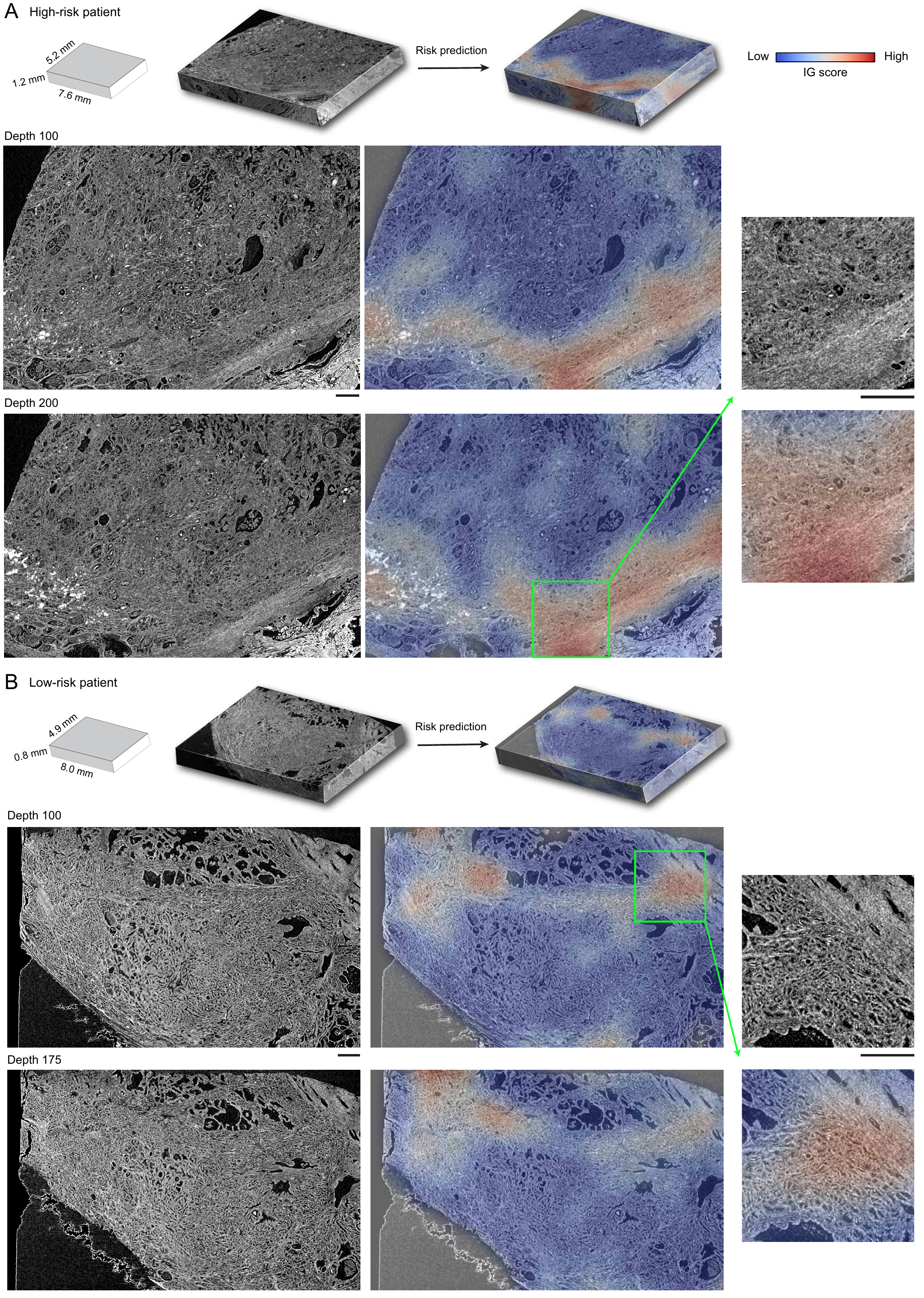}
\end{figure}

\newpage
\captionof{figure}{\textbf{Integrated gradient (IG) heatmaps for the microcomputed tomography (microCT) cohort} The integrated gradient scores are assigned to each patch with high IG (low IG) patch indicating that patch contributes to unfavorable (favorable) prognosis. (A) In this high-risk sample, high IG values are localized in areas with the smallest and densest cancerous glands, especially when they are in or adjacent to the capsule of the prostate, as well as dense stroma that resembles the prostate capsule. (B) Similar to the high-risk case, high IG regions in this low-risk sample correspond to areas with small, dense cancerous glands and dense stroma. The juxtaposition of these two morphologies has particularly high IG values. All scale bars are $500 \mu m$. The heatmaps can also be visualized in our interactive demo.}
\label{fig:ct_heatmaps}

\newpage
\begin{figure}[!ht]
    \centering
    \includegraphics[width=0.93\textwidth]{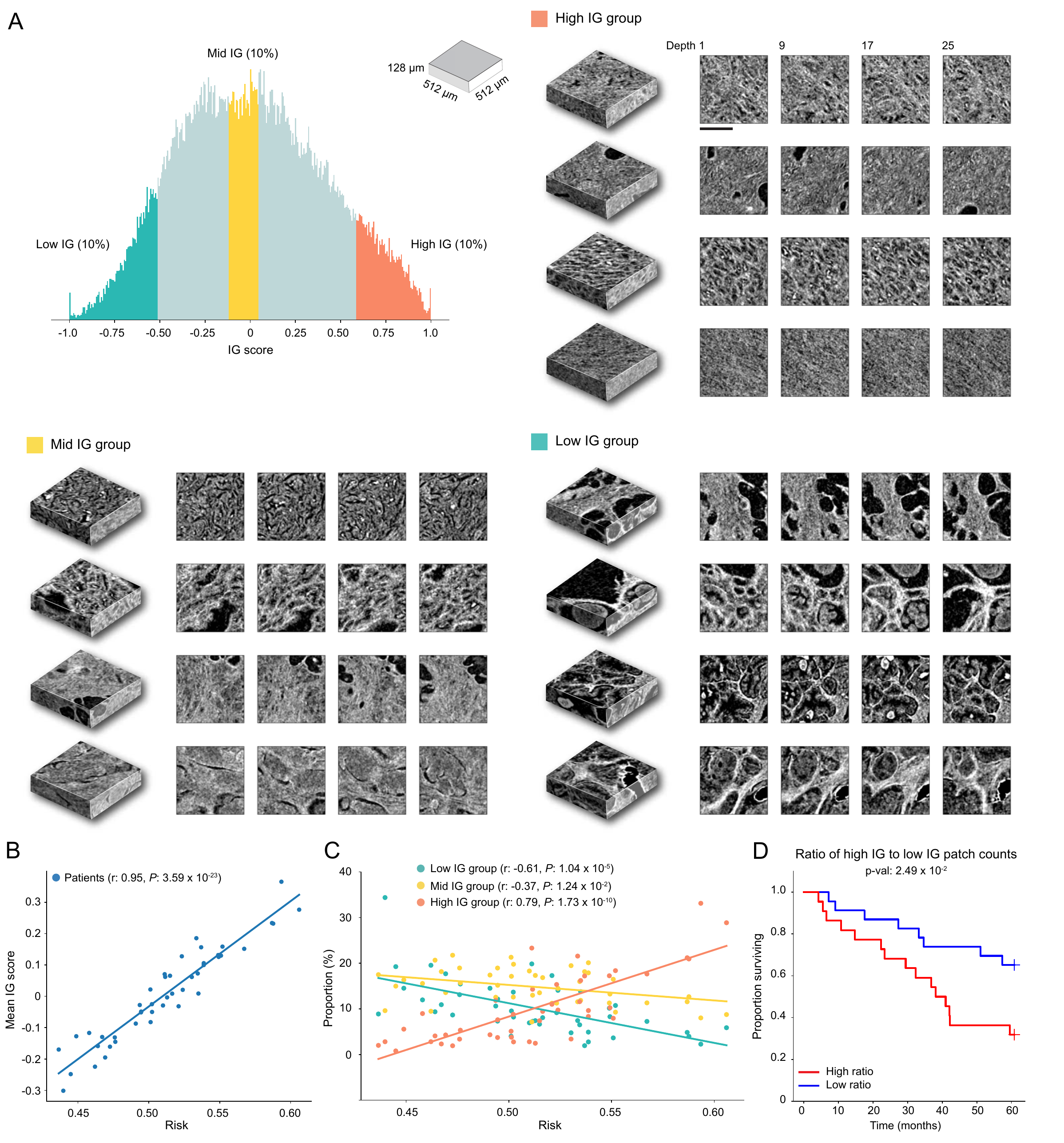}
    \caption{\textbf{Integrated gradient analysis for microcomputed tomography (microCT) dataset} (A) The high IG cluster consists of patches with infiltrative carcinoma that most closely resembles Gleason pattern 4; however, the lower resolution and lack of H\&E staining make definitive grading effectively impossible by visual inspection of the microCT images alone. In the middle IG cluster, most patches contain infiltrating carcinoma that resembles Gleason patterns 3 and 4. The low IG cluster consists almost mostly of patches containing benign prostatic tissue with occasional foci of infiltrative carcinoma that resembles Gleason pattern 3.
    (B) Scatter plot of the normalized IG patch scores averaged within each sample as a function of predicted risk (the predicted probability for the high-risk group). (C) The scatter plot of the proportion of the number of high, middle, and low IG group patches in each sample as a function of predicted risk, which shows that the sample with higher predicted risk has a larger (smaller) fraction of high (low) IG patches. (D) Kaplan-Meier curve for the cohort stratified (50\%) by the ratio of the number of patches in the high and low IG group. The good stratification performance suggests that the extent to which prognostic morphologies manifest in each sample are also important. The scale bar is $250 \mu m$.}
    \label{fig:ct_ig_ablation}
\end{figure}


\newpage
\begin{figure}[!ht]
    \centering
    \includegraphics[width=0.98\textwidth]{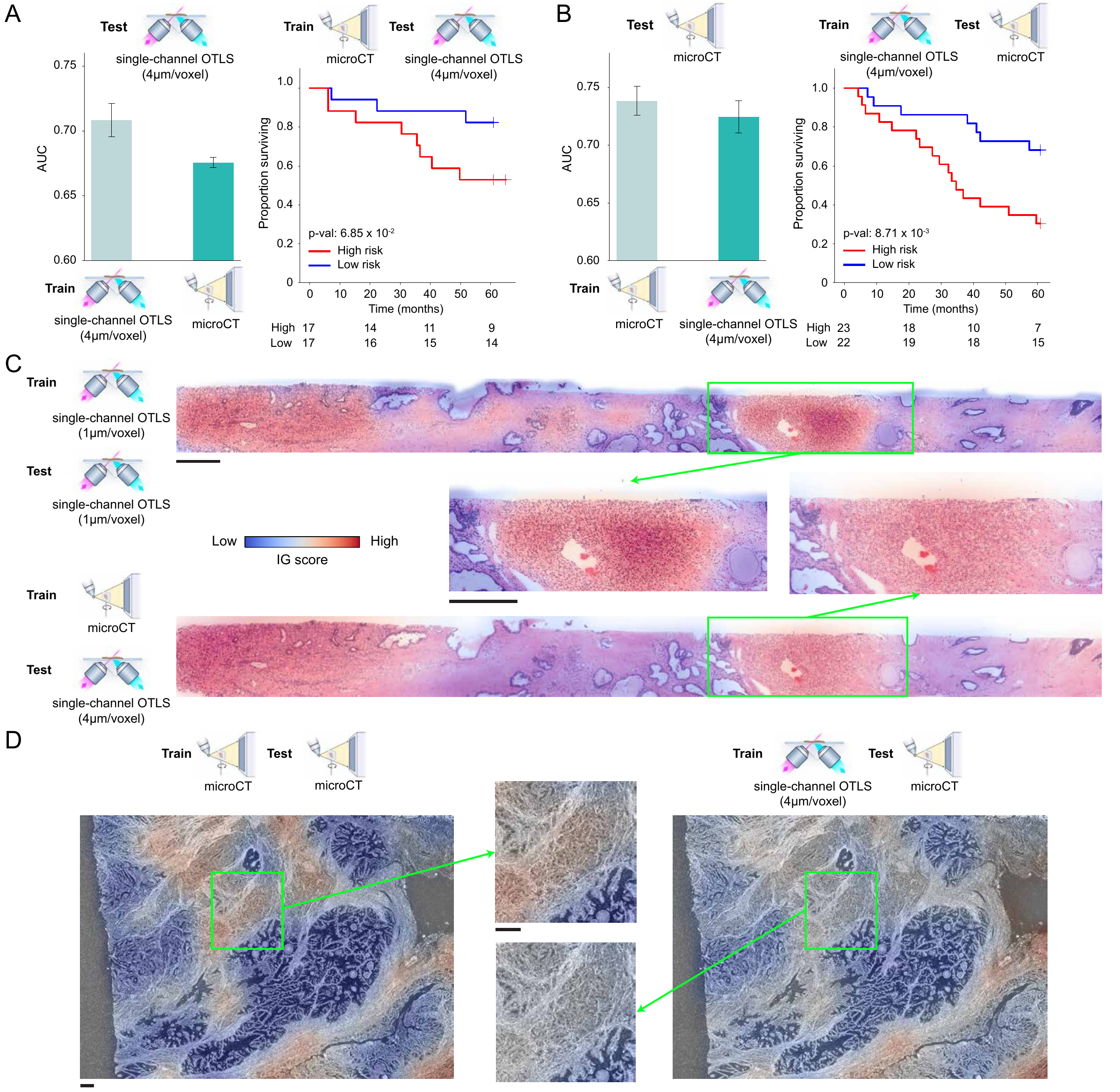}
    \caption{\textbf{Cross-modal evaluation between open-top light-sheet microscopy (OTLS) and microcomputed tomography (microCT) datasets.} We perform a cross-modal experiment by training a network with the whole block cuboid setting on one cohort and testing on the other to assess whether the network learns generalizable prostate cancer prognostic morphologies. To match the $4\mu m$/voxel resolution and single-channel characteristics of the microCT dataset, the OTLS dataset is downsampled by a factor of $4$, and only the nuclear channel is retained. (A) Test AUC for the OTLS cohort with MAMBA trained on OTLS or microCT cohorts, and the cross-modal Kaplan-Meier curve for cohort stratification of high and low-risk groups. (B) Identical analyses to (A), but tested on microCT with MAMBA trained on microCT or OTLS cohorts. The drop in test AUC for both experiments can be attributed to significantly different imaging protocols between the two datasets. (C-D) Integrated gradient (IG) heatmaps for cross-modal experiments. Despite the difference in train and test modalities, MAMBA identifies poorly-differentiated glands (C) and infiltrative carcinoma (D) as unfavorable prognostic morphologies, concurring with IG heatmaps from the original same-modality setting. These results suggest that despite the challenging nature of the cross-modal adaptation, MAMBA identifies important prognostic morphologies robust to different image pipelines. All scale bars are $250 \mu m$.}
    \label{fig:cross_modality}
\end{figure}

\newpage
\begin{figure}[!ht]
    \centering
    \includegraphics[width=\textwidth]{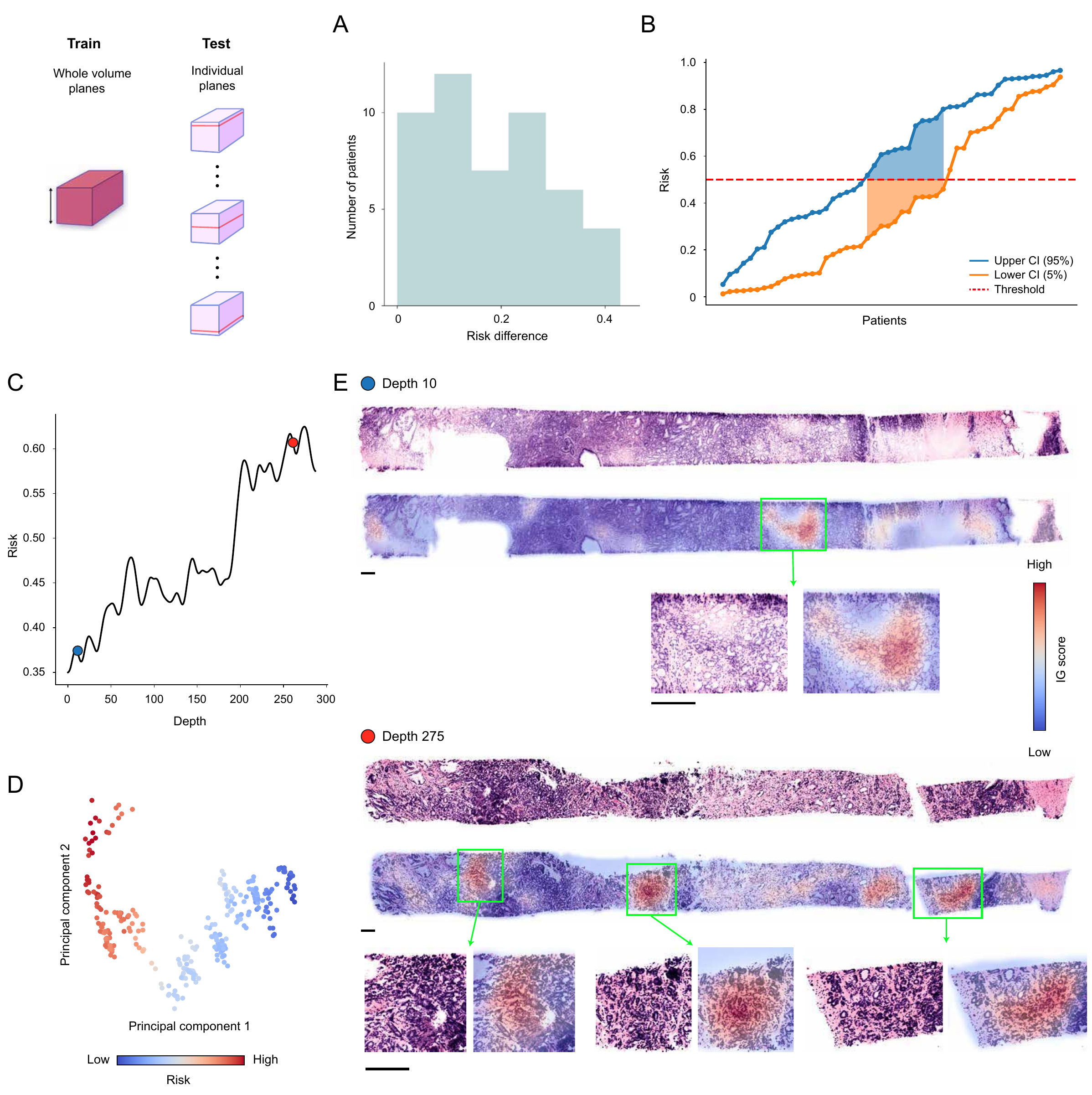}
    \caption{\textbf{Plane variability analysis for open-top light-sheet microscopy (OTLS) dataset} We use residual CNN (ResNet50) feature encoder to train MAMBA on all planes of whole volume and predict risk (the probability for the high-risk group) on individual planes. (A) Given the plane-level predicted risks for each sample, the difference between lower 5\% and upper 95\% value is computed (risk difference). The histogram shows that the difference is non-negligible for some patients, indicating heterogeneity within the tissue. (B) An arbitrary risk decision threshold (\textit{e.g.}, 0.5) falls within the 90\% risk interval for several patients, for whom the risk group can change depending on the plane chosen for prognosis. (C) Plane-level predicted risk, which fluctuates from low-risk to high-risk, as a function of depth within the volume for a patient. (D) Principal component feature space for attention-aggregated plane-level features for the sample. The separation into two clusters along the risk group reflects the risk variation observed in (C). (E) Morphological analysis of the low-risk (depth 10) and high-risk plane (depth 275) agrees with the prediction, with the higher-risk plane containing a larger proliferation of tumor resembling Gleason pattern 4 morphology than the lower-risk plane.}
    \label{fig:otls_variability_2D}
\end{figure}

\newpage
\begin{figure}[!ht]
    \centering
    \includegraphics[width=\textwidth]{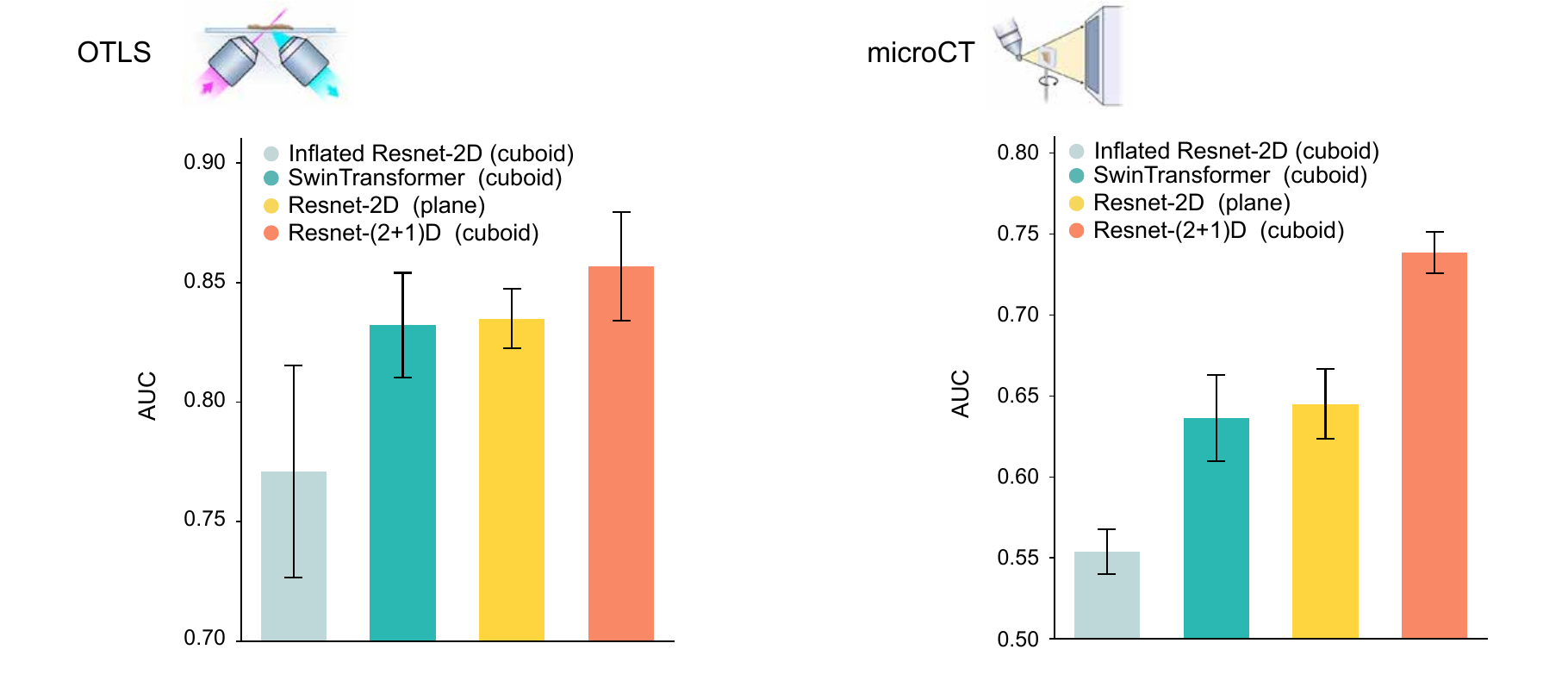}
    \caption{\textbf{Comparsion between different feature encoders} We perform ablation studies on different feature encoders for OTLS and microCT datasets. MAMBA relies on transfer learning to extract representative and compressed features from 2D patches and 3D patches. MAMBA provides access to a diverse range of feature encoders for users to choose from. The results demonstrate that different feature encoders utilizing the whole volume lead to varying performance levels, with the spatiotemporal CNN\cite{tran2018closer} (\textit{i.e.}, ResNet-(2+1)D) used for our study, performing the best for both OTLS and microCT datasets.}
    \label{fig:extractor_ablation}
\end{figure}

\clearpage
\setcounter{table}{0}
\renewcommand{\tablename}{Extended Data Table}

\begin{table}[ht!]
\centering
\caption{\textbf{Dataset summary for OTLS and microCT cohort} Gleason grades are based on pathology reports.}
\begin{tabular}{l  c  c } 
\hline
Statistic & \textbf{OTLS} & \textbf{microCT}\\
\hline
Number of Patients & 50 & 45 \\
Number of tissue specimens & 118 (1 - 4 per patient) & 45 (1 per patient)\\
Age (years) at diagnosis &&\\
\tabindent $<60$ & 27 (54\%) & 18 (40\%)\\
\tabindent $>60$ & 23 (46\%) & 27 (60\%)\\
Biochemical recurrence (BCR) &&\\
\tabindent BCR within 5 years & 25 (50\%) & 23 (51\%)\\
\tabindent No BCR within 5 years & 25 (50\%) & 22 (49\%)\\
Gleason grade &&\\
\tabindent 3+3 & 53 (45\%) & 7 (16\%) \\
\tabindent 3+4 & 36 (31\%) & 23 (51\%) \\
\tabindent 4+3 & 23 (19\%) & 11 (24\%) \\
\tabindent 4+4 & 6 (5\%) & 4 (9\%) \\
\hline
\end{tabular}
\label{table:data_summary}
\end{table}

\newpage
\begin{table}[ht!]
    \caption{\textbf{MAMBA computational complexity}}
    \begin{subtable}{\linewidth}
    \centering
    \caption{\textbf{Dataset summary}}
    \begin{tabular}{|l || c c | c c|} 
    \hline
    & \multicolumn{2}{c|}{\bfseries OTLS} & \multicolumn{2}{c|}{\bfseries microCT}\\
    & planes (2D) & cuboids (3D) & planes (2D) & cuboids (3D)\\
    \hline
    Dimensions (voxels) & \multicolumn{2}{c|}{$320\times 520 \times 9,500\,\, (1.58\times 10^9)$} & \multicolumn{2}{c|}{ $1,300 \times 3,200\times 1,920\,\, (7.99\times 10^9)$}\\
    Patch size (voxels) & $128 \times 128 \times 1$ & $128 \times 128  \times 64$ & $128 \times 128 \times 1$ & $128 \times 128 \times 32$  \\
    Number of patches & $3.7\, (\pm\, 1.2)\times 10^4$ & $1.2\, (\pm\, 0.3)\times 10^3$ & $8.0\,(\pm\,2.9) \times 10^4$ & $2.5\,(\pm\,0.9) \times 10^3$ \\ 
    \hline
    \end{tabular}
    
    \vspace{0.6cm}

    \caption{\textbf{Average runtime for MAMBA processing} \tablefootnote{All rutimes are in seconds. For training and inference, runtimes are per epoch and sample. Additional details for runtime computation are in Methods.}}
    \begin{tabular}{|l||c|c|c|c|}
    \hline
    & Encoding (s / $10^4$ patches)& Encoding (s / sample)  & Training (s) & Inference (s) \\
    \hline
    planes (2D) & $3.15 \times 10^0$ & $2.29\times 10^1$ & $9.8 \times 10^{-1}$ & $3.0 \times 10^{-1}$\\
    cuboids (3D) & $1.17\times 10^2$  & $2.66\times 10^1$ & $2.5 \times 10^{-2}$ & $1.1 \times 10^{-2} $\\
    \hline
    \end{tabular}
    \end{subtable}
    \label{table:data}
\end{table}

\end{document}